# Graphene quantum dots prevent α-synucleinopathy in Parkinson's disease

Donghoon Kim[1,2,*], Je Min Yoo[8,*], Heehong Hwang[1,7], Junghee Lee[10], Su Hyun Lee[1,2], Seung Pil Yun[1,2,6], Myung Jin Park[8], MinJun Lee[8], Seulah Choi[1], Sang Ho Kwon[1], Saebom Lee[1,2], Seung-Hwan Kwon[1,2], Sangjune Kim[1,2], Yong Joo Park[3], Misaki Kinoshita[11], Young-Ho Lee[11], Seokmin Shin[8], Seung R. Paik[9], Sung Joong Lee[7], Seulki Lee[3,4], Byung Hee Hong[8,†] and Han Seok Ko[1,2,5,6,†]

[1]Neuroregeneration and Stem Cell Programs, Institute for Cell Engineering, The Johns Hopkins University School of Medicine, Baltimore, Maryland, United States of America

[2]Department of Neurology, The Johns Hopkins University School of Medicine, Baltimore, Maryland, United States of America

[3]The Russell H. Morgan Department of Radiology and Radiological Sciences, Johns Hopkins University School of Medicine, Baltimore, Maryland, United States of America

[4]The Centre for Nanomedicine at the Wilmer Eye Institute, Johns Hopkins University School of Medicine, Baltimore, Maryland, United States of America

[5]Diana Helis Henry Medical Research Foundation, New Orleans, Louisiana, United States of America

[6]Adrienne Helis Malvin Medical Research Foundation, New Orleans, Louisiana, United States of America

[7]Department of Neuroscience and Physiology, Interdisciplinary Program in Neuroscience, Dental Research Institute, Seoul National University, Seoul 08826, Republic of Korea

[8]Department of Chemistry, College of Natural Science, Seoul National University, Seoul 08826, Republic of Korea

[9]School of Chemical and Biological Engineering, College of Engineering, Seoul National University, Seoul 08826, Republic of Korea

[10]Inter-University Semiconductor Research Centre, Seoul National University, Seoul 08826, Republic of Korea

[11]Institute for Protein Research, Osaka University, Yamadaoka 3-2, Suita, Japan

[*, †] *These authors contributed equally to this work.*

Correspondence

Han Seok Ko, Ph.D.
Neuroregeneration and Stem Cell Programs

Institute for Cell Engineering
Johns Hopkins University School of Medicine
733 North Broadway, Suite 731
Baltimore, MD 21205
Email: hko3@jhmi.edu

Or

Byung Hee Hong, Ph.D.
Department of Chemistry
College of Natural Science
Seoul National University
Seoul 440-746, South Korea
Email: byunghee@snu.ac.kr

While emerging evidence indicates that the pathogenesis of Parkinson's disease is strongly correlated to the accumulation[1,2] and transmission[3,4] of α-synuclein (α-syn) aggregates in the midbrain, no anti-aggregation agents have been successful at treating the disease in the clinic. Here we show that graphene quantum dots (GQDs) inhibit fibrillization of α-syn and interact directly with mature fibrils, triggering their disaggregation. Moreover, GQDs can rescue neuronal death and synaptic loss, reduce Lewy body and Lewy neurite formation, ameliorate mitochondrial dysfunctions, and prevent neuron-to-neuron transmission of α-syn pathology provoked by α-syn preformed fibrils (PFFs)[5,6]. We observe that *in vivo*, GQDs penetrate the blood-brain barrier (BBB) and protect against dopamine neuron loss induced by α-syn PFFs, Lewy body/Lewy neurite pathology, and behavioural deficits (Supplementary Fig. 1).

Following the synthesis and analysis of GQDs (Supplementary Fig. 2a-c,e), the potential role of GQDs in inhibiting α-syn fibrillization and disaggregating fibrils was preliminarily investigated (Fig. 1a). In the absence of GQDs, α-syn monomers are assembled into mature fibrils as assessed by thioflavin T (ThT) fluorescence (Fig. 1b), turbidity assays (Fig. 1c), and transmission electron microscopy (TEM) analysis (Fig. 1d). In contrast, the same assessments display that GQDs predominantly inhibit α-syn fibrillization. Moreover, GQDs induce dissociation of α-syn fibrils into short fragments (Fig. 1e-i, supplementary Fig. 3a), where the average length of the fragments shortens from 1 μm to 235 and 70 nm after 6 and 24 hours, respectively (Fig. 1g,i). The average number of the shortened fragments is increased during the first 24 hours of incubation, suggesting that the fibrils are dissociated from the inner parts (Fig. 1h). However, the number of fragments starts to decrease at day 3 and is no longer detectable at day 7, indicating complete dissociation of fibrils in the course of time (Fig. 1i). Time-dependent atomic force microscopy (AFM) images show the same dissociation process, where GQDs and fibrils can be differentiated by their distinguishing height profiles (Supplementary Fig. 3b). Dot-blot assay and blue native polyacrylamide gel electrophoresis (BN-PAGE) analysis also reveal that incubation with GQDs results in gradual reduction of fibrils in a time-dependent manner (Fig. 1j,k). Similar effects of GQDs are observed for sonicated α-syn PFFs, which produced higher population of monomers as a function of the incubation period (Supplementary Fig. 4a-c).

To visualise the interaction between GQDs and α-syn fibrils, GQDs were functionalised with PEGylated biotin (Supplementary Fig. 2d-g). As detected with streptavidin tagged ultra-small gold particles using TEM, the direct binding between GQDs and α-syn fibrils was visualised (Fig. 2a,b). Two-dimensional $^{1}H$-$^{15}N$ hetero-nuclear single-quantum coherence correlation (HSQC) NMR spectroscopy was utilised to further analyse the interaction site of $^{15}N$-labelled α-syn[7,8] for GQDs and their molecular interactions at the residue level in solution. The co-incubation of GQDs produced many residues with large chemical shifts and entirely disappeared residues (Supplementary Fig. 5). The disappeared residues are centred on the N-terminal region, implying that the initial binding between GQDs and α-syn is largely driven by the charge interaction between GQDs' negatively charged carboxyl groups (Supplementary Fig. 2e) and the positively charged region of α-syn (Fig. 2c).

To better understand and elucidate the mechanism of α-syn fibril dissociation by GQDs, 200 ns molecular dynamics (MD) simulation was performed. The structure of α-syn fibril was adopted from the recently reported ssNMR (solid-state nuclear magnetic resonance) structure of pathologic human α-syn[9]. To facilitate the simulation process, only the non-amyloid-β component (NAC) domain (residue 71 to 82) was taken as it is essential for α-syn fibrillization[10]. Following the instantaneous binding between GQDs and the N-terminal cross-β part of α-syn at 1 ns, the β-sheet structure of the outer monomer is completely destroyed after 50 ns; its C-terminal part is released from the core and interacts with the opposite plane of GQDs (Fig. 2d). The time-dependent secondary structure plot, calculated by the dictionary of secondary structure of proteins (DSSP) algorithm, also presents decrease in the β-sheet component of the outer monomer from 50 ns onwards, indicating a critical structural disruption in the fibril (Fig. 2f). For further analysis, the changes in the root-mean-square deviation (RMSD) of atomic positions, solvent accessible surface area (SASA), and the interaction energies of the fibril were plotted against time (Fig. 2e). While massive changes in the RMSD and SASA values are observed after 50 ns, the changes in the total potential energy ($\Delta U_{tot}$) show a slight, yet continuous decline. It is confirmed that the major stabilising/dissociation force is attributed to strong hydrophobic interactions, as the changes in the van der Waals energy ($\Delta E_{van}$) show a decline and the electrostatic energy ($\Delta E_{elec}$) remains steady. The results are consistent with the 65 ns snapshot, which shows strong hydrophobic interactions between GQDs' basal plane and valine residues (Fig. 2d). To further analyse the changes in the secondary structure of α-syn fibrils, circular dichroism (CD) spectra were obtained and the fractional secondary structures were analysed using the algorithm of CONTIN/LL[11,12] (Fig. 2g,h). After 7 days, changes in the secondary structure were observed, where the β-sheet component decreased from 53.3±3.5% to 29.8±3.4% and the α-helix / random coil components increased from 4.2±1.2% to 19.8±1.5% and 20.1±5.7% to 24.6±3.6%, respectively. Collectively, the interaction between GQDs and α-syn is initiated by the charge interactions, and fibril dissociation is driven chiefly by hydrophobic interactions, accompanied by the structural changes.

Based on these results, GQDs' potential role in preventing α-syn PFFs-induced pathology was explored in primary neurons. α-syn PFFs treatment leads to critical cell death as assessed by various cell viability assays including terminal deoxynucleotidyl transferase dUTP nick end labelling (TUNEL) (Fig. 3a and Supplementary Fig. 6a), alamarBlue (Fig. 3b), lactate dehydrogenase (LDH) (Fig. 3c), and neurite outgrowth assays (Supplementary Fig. 6b-d). In contrast, the presence of GQDs reduces α-syn PFFs-induced cell toxicity in the same assessments. In addition, treatment of α-syn PFFs leads to reduction in synaptic proteins such as SNAP25 and VAMP2, suggesting severe dysfunction in neuronal networks[5]. Importantly, GQDs restore the reduced synaptic protein levels provoked by α-syn PFFs (Supplementary Fig. 6e-g). Mitochondrial dysfunction is another major pathological hallmark during the progression of PD[13]. The effect of GQDs on mitochondrial dysfunction and subsequent cellular respiration was thus investigated by performing oxidative stress marker 8-hydroxyguanosine (8-OHG) staining, mitochondrial complex I activity assay, MitoTracker staining, TEM analysis, and seahorse assay (Supplementary Fig.7). While the treatment of α-syn PFFs causes mitochondrial damages characterised by shrinkage, decreased oxygen consumption rate, and elevated reactive oxygen species (ROS) levels in neurons, the addition of GQDs ameliorates these adverse effects.

Next, the role of GQDs on α-syn PFFs-induced LBs/LNs-like pathology was evaluated. After 7 days of α-syn PFFs treatment, accumulation of α-syn in the sodium dodecyl sulphate (SDS)-insoluble fraction was analysed by Western blot, which shows decreased phosphorylated α-syn (p-α-syn) accumulation with GQDs (Fig. 3d,e). In addition, p-α-syn immunoreactivity is increased in primary neurons by α-syn PFFs, whereas it is barely detectable in primary neurons treated with GQDs as assessed by α-syn phosphoserine 129 immunostaining (Fig. 3f,g). To investigate the effect of treatment time on α-syn pathology, GQDs were treated simultaneously with and 3 days pre- / post α-syn PFFs additions. While the toxicity amelioration and inhibition effects on p-α-syn accumulation were slightly weaker when GQDs were treated 3 days after α-syn PFFs injection, the collective results practically show consistent effects (Supplementary Fig.

8a-f). To verify the cellular localisation of GQDs' inhibitory effects, GQDs, α-syn PFFs, and the lysosome were distinctively labelled with fluorescent probes for live imaging (Supplementary Fig. 8g,h). As shown from the time-course snapshots, the fluorescence signals for α-syn PFFs and GQDs are co-localised within the lysosome, where the signal for α-syn PFFs gradually decreases and that of GQDs increases with time. It can be presumed that both GQDs and α-syn PFFs are endocytosed, and the disaggregation of fibrillized α-syn takes place in the lysosome by surrounding GQDs.

One of the most key features of the α-syn PFFs neuron model is neuron-to-neuron transmission of pathologic α-syn aggregates[5]. To examine the effect of GQDs, a triple-compartment microfluidic culture device was employed, where transmission between neurons takes place sequentially from C1 to C3 (Fig. 3h). At 14 days post-α-syn PFFs treatment, the levels of p-α-syn were visualised by immunostaining in all test chambers. The levels of p-α-syn immunoreactivity are greatly reduced in C2 and C3 in the presence of GQDs in C1, which suggests that GQDs prevent the seeds of endogenous α-syn from forming aggregates (Fig. 3i,j). Also, the levels of p-α-syn immunoreactivity are lowered in C2 and C3 when GQDs are added to C2; GQDs block the transmission of pathologic α-syn to the neighbouring neurons.

To address whether GQDs possess neuroprotective effects against α-syn PFFs-induced transmission and toxicity *in vivo*, α-syn PFFs were stereotaxically injected into the striatum of wild type (WT) mice and GQDs were biweekly administered via intraperitoneal (i.p.) injection (Fig. 4a). Foremost, we sought to verify the blood-brain barrier (BBB) permeability of GQDs by employing an *in vitro* BBB model[14] (Supplementary Fig. 9a,b). The formation and functionality of the BBB model were subsequently confirmed by measuring the transepithelial electrical resistance (TEER) and dextran fluorescence intensity (Supplementary Fig. 9c,d). The *in vitro* permeability of GQDs was then ascertained by measuring the fluorescence intensity on the brain side. Intriguingly, the fluorescence signal gradually increased with time, and showed 100% permeability after 24 hours (Supplementary Fig. 9e). To corroborate the penetration process visually, GQDs-biotin was utilised and linked to streptavidin-quantum dots for live fluorescence imaging. Prior to the study, the functional behaviours of GQDs-biotin were assessed, which exhibited similar BBB penetration rates and fibril dissociation effects as those of pristine GQDs (Supplementary Fig. 9e-g). Live fluorescence imaging shows that GQDs-biotin complex is localised in the lysosome of both BMEC and astrocytes, and disappears through the exosome (Supplementary Fig. 9h,i). It suggests that GQDs are endocytosed into BMEC on the blood side and released, which are subsequently endocytosed by astrocytes on the brain side and released through the exosome. *In vivo* permeability of the BBB was studied by employing GQDs-biotin for immunohistochemical analysis of the brain. Considerable amount of GQDs-biotin was detected in the entire central nervous system (CNS) region including the olfactory bulb, neocortex, midbrain, and cerebellum after i.p. injection, indicating that GQDs have the ability to penetrate the BBB *in vivo* as well (Supplementary Fig. 9j-n).

At 180 days post-α-syn PFFs injection, decreased tyrosine hydroxylase (TH)- and Nissl-positive neurons in the substantia nigra (SN) are observed in WT mice. On the other hand, mice administered with GQDs are protected against α-syn PFFs-induced loss of dopaminergic neurons (Fig. 4b,c). Furthermore, α-syn PFFs provoke a loss of TH-positive striatal fibre neurons, while GQDs prevent the loss in the striatum as well (Fig. 4d,e). Finally, changes in the behavioural deficits were assessed by the cylinder and pole tests. Mice with GQDs injection exhibited alleviated motor deficit, showing balanced use of both forepaws and decreased pole descending time, respectively in the cylinder and pole tests (Fig. 4f,g).

The levels of p-α-syn, a marker of LBs/LNs were subsequently visualised in the striatum and SN of α-syn PFFs-injected mice. While stereotaxically injected α-syn PFFs provoke accumulation of p-α-syn in the striatum and SN, the administration of GQDs reduces the p-α-syn levels (Fig. 4h,i). The injection of α-syn PFFs in the striatum also leads to propagation of α-syn aggregates throughout the CNS region, whereas GQDs inhibits the transmission of pathological α-syn (Fig. 4j). In addition, GQDs injection ameliorates α-syn PFFs-induced gliosis in the SN accompanying the decreased microglia density and glial fibrillary acidic protein (GFAP) levels in astrocytes (Supplementary Fig.10). To further corroborate GQDs' therapeutic potential against transgenic *in vivo* PD model, 6-month human A53T α-syn transgenic mice[15] were biweekly administered with GQDs for 4 months (Supplementary Fig. 11a). Similar to the α-syn PFFs-induced model, GQDs administration reduces the p-α-syn levels in the CNS region, (Supplementary Fig. 11b-d) as well as the microglia density and GFAP levels in astrocytes in the transgenic model (Supplementary Fig. 11h-k). GQDs treatment also reduces the α-syn aggregates induced by overexpression of human A53T α-syn in HEK293 cells[16] (Supplementary Fig. 11l,m). GQDs alleviate the behavioural defects in the hA53T α-syn Tg mice as well, monitored by the pole and clasping tests (Supplementary Fig. 11e-g). It must be noted that there is no loss of dopamine neurons, glial cells activation, behavioural abnormalities, and organ damages in mice with 6 months of prolonged GQDs injection, demonstrating that GQDs manifest no appreciable long-term *in vitro* and *in vivo* toxicity and can be cleared from the body and excreted into urine (Supplementary Fig. 12).

A few previous reports have discussed graphene oxides' (GOs) and GQDs potential therapeutic role against Alzheimer's disease (AD) by inhibiting the fibrillization of beta-amyloid (Aβ) peptides[17-19]. In addition, computational evidence suggests hydrophobic graphene sheet could cause destruction of amyloid fibrils[20]. To compare the therapeutic efficacy of GQDs with the previous studies, nano-sized GOs (nano-GOs) ranging 5 ~ 20 nm and reduced GQDs (rGQDs) were prepared (Supplementary Fig. 13a,b). As shown from the FT-IR spectra, pristine GQDs exhibit predominant carboxyl groups, while nano-GOs and rGQDs show much less carboxyl group: aromatic carbon double bond ratios. The difference in the carboxyl functional group contents was directly reflected in cell viability assays, where nano-GOs and rGQDs show serious *in vitro* toxicities unlike GQDs (Supplementary Fig. 13c,d). Moreover, nano-GOs and rGQDs exhibit considerably weaker fibril-dissociation effects than GQDs (Supplementary Fig. 13e-g). This can be presumably attributed to the presence of less carboxyl functional groups as well, as we have verified from the $^{1}H$-$^{15}N$ HSQC analysis that the positively charged N-terminal region is the initial, yet the chief binding site with GQDs. It must be also noted that the BBB permeability of nano-GOs is much lower than that of GQDs and rGQDs, suggesting that the size may be a decisive factor for the BBB permeability of graphene-based nanoparticles (Supplementary Fig. 13h). Collectively, our results show that GQDs are the optimal therapeutic candidate for anti-PD and related α-synucleinopathies therapy with no appreciable *in vitro* and long-term *in vivo* toxicity, respectable fibril dissociation effects, and the ability to pass through the BBB.

In order to open new venues in clinical drug development against PD, a candidate desirably features outstanding anti-aggregation and dissociation properties towards α-syn aggregates without severe toxicity. GQDs are found to bind to α-syn fibrils, inhibit transmission, and possess unique neuroprotective effects against the neuropathological α-syn aggregates/fibrils in both *in vitro* and *in vivo* models. It is expected that GQDs-based drugs with appropriate modifications might provide a clue to support the development of new therapeutic agents for abnormal protein aggregation related neurological disorders including PD.

## Acknowledgements

This work was supported by NRF (National Research Foundation of Korea) grant funded by the Korean government (NRF-2014H1A2A1016534-Global Ph.D. Fellowship Program, NRF-2011-357-C00119) and grants from the NIH/NINDS NS38377 Morris K. Udall Parkinson's Disease Research Center, NIH/NINDS NS082205, and NIH/NINDS NS098006. This work was made possible by support from the Johns Hopkins Medicine Discovery Fund. The authors acknowledge the joint participation by the Adrienne Helis Malvin Medical Research Foundation and the Diana Helis Henry Medical Research Foundation through its direct engagement in the continuous active conduct of medical research in conjunction with The Johns Hopkins Hospital and the Johns Hopkins University School of Medicine and the Foundation's Parkinson's Disease Program M-2014, H-1, H-2013. We would like to extend our sincere gratitude to Professor Hyukjin Lee of Ewha Womans University for discussions and helpful advices.

## Additional Information

Supplementary information is available in the online version of the paper. Reprints and permission information is available online at www.nature.com/reprints. Correspondence and requests for materials should be addressed to H.S.K. or B.H.H.

## Methods

**Preparation of GQDs.** GQDs were synthesised by putting 0.9 g of carbon fibres (Carbon Make) in a mixture of strong acid (300 ml sulphuric acid & 100 ml nitric acid; Samchun Chemical) at 80 $^{o}$C for 24 hours. After acid removal, the solution was vacuum-filtered with porous inorganic membrane filter (Cat#: 6809-5002; Whatman-Anodisc 47; GE Healthcare) to discard large particles. The solution was then subjected to rotary evaporation to yield the final product in powder form.

**FT-IR measurements.** The samples prepared through the conventional KBr pellet method (Number of scan: 32; Resolution: 4; Wavenumber range: 4000 – 40 $cm^{-1}$). The analysis was performed with a Nicolet 6700 FT-IR spectrometer (Thermo Scientific).

**ThT and Turbidity assays.** α-syn fibrils were measured with both turbidity and ThT assays. For ThT assay, 50 μl of each sample was centrifuged for 30 mins at 16,000 × g. The pellet was resuspended in 200 μl of 25 μM ThT (Cat#: T3516; Sigma-Aldrich) in 10 mM glycine buffer (pH 9.0). ThT fluorescence was measured at 482 nm (excitation at 440 nm) by a fluorescence spectrophotometer. For turbidity assay, α-syn fibrils were diluted (1/10) with PBS. The diluted fibrils were transferred to Corning® 96-well plates and the absorbance intensity at 360 nm was measured to assess the turbidity of each sample by microplate multi-reader.

**TEM imaging.** For fibrils, samples were adsorbed to glow discharged 400 mesh carbon coated copper grids (EMS) for 2 mins. The grids were transferred through three drops of 50 mM Tris-HCl (pH 7.4) rinse, then floated upon two consecutive drops of 0.75% uranyl formate, each for 30 secs. Stain was either aspirated or blotted off with #1 Whatman filter paper triangles. Grids were allowed to dry before imaging by a Phillips CM 120 TEM operating at 80 kV. Images were captured and digitised with an ER-80 CCD (8 megapixel) by AMT. For neurons, primary cortical neurons were plated at a density of 100,000 cells/well onto the 35 mm dish coated with poly-*D*-lysine. Neurons were treated with 1 μg/ml of PFFs with or without 1 μg/ml of GQDs at days in vitro (DIV) 10. After 7-day treatment, neurons were washed with PBS containing 1% sodium nitrite (pH 7.4), fixed with fixative consisting of 3% (vol/vol) paraformaldehyde (PFA), 1.5% (vol/vol) glutaraldehyde, 100 mM cacodylate, and 2.5% (vol/vol) sucrose (pH 7.4), and post-fixed for 1 hour. Images were collected on a Philips EM 410 TEM installed with a Soft Imaging System Megaview III digital camera.

**Dot-blot assays.** Samples were loaded onto the pre-wetted nitrocellulose membrane (0.45 μm pore-sized) using the Bio-Dot microfiltration apparatus (Cat#: 1706545; Bio-Rad) and allowed to filter through the membrane under mild vacuum. After washing each sample with Tris-buffered saline, samples were blocked with 5% non-fat dry milk in Tris-buffered saline containing tween-20. Samples were incubated with conformation-specific anti-α-syn filament antibody (Cat#: ab209538; 1:1,000; abcam) at 4 $^{o}$C overnight, followed by HRP-conjugated rabbit secondary antibodies (GE Healthcare) for 1 hour at RT. Blots were visualised by ECL solution and analysed with ImageJ software (http://rsb.info.nih.gov/ij/,NIH).

**BN-PAGE and SDS-PAGE.** For BN-PAGE, α-syn fibrils and α-syn PFFs were prepared using NativePAGE™ sample prep kit (Cat#: BN2008; Life technologies) and run on NativePAGE™ Novex 4-16% Bis-Tris protein gels (Cat#: BN1002Box; Life technologies) at 200 V for 90 mins. The cathode buffer contained 50 mM tricine, 15 mM Bis-Tris, 0.02% Brilliant Blue G (pH 7.0), and the anode buffer was consisted of 50 mM Bis-Tris (pH 7.0). Gels were stained using the SilverQuest™ silver staining kit (Cat#: LC6070; Life technologies), following the manufacturer's instructions. For SDS-PAGE, 10 DIV cortical neurons were treated with α-syn PFFs (5 μg/ml) in the presence and absence of GQDs (5 μg/ml) for 7 days. Soluble proteins of neurons were prepared in 1% TX-100 in PBS and protease and phosphatase inhibitor cocktail (Cat#: PPC1010; Sigma-Aldrich) at 4 $^{o}$C. Lysates were sonicated and centrifuged at 12,000 × g for 30 mins at 4 $^{o}$C. The pellet was washed several times and suspended in 2% SDS in PBS for insoluble protein preparation. 2× Laemmli sample buffer (Cat#: 1610737; Bio-rad) was utilised to dilute the lysates. Subsequently, 20 μg of proteins were loaded onto the wells of Novex™ 8-16% Tris-Glycine Gel (Cat#: XP08160BOX; Life technologies) and electrophoresis was performed at 130 V for 85 mins. The proteins were then transferred onto nitrocellulose membrane, blocked with 5% non-fat dry milk in TBS with 0.1% Tween-20 for 1 hour and incubated at 4 $^{o}$C overnight with anti-pS129-α-syn (Cat#: ab59264; 1:1,000, abcam), SNAP25 (Cat#: 111-002, 1:2,000, Synaptic Systems) or VAMP2 (Cat#: ab3347; 1:1,000, abcam) antibodies, followed by HRP-conjugated rabbit or mouse secondary antibodies (GE Healthcare) for 1 hour at RT. Blots were visualised by ECL solution and analysed with ImageJ software (http://rsb.info.nih.gov/ij/,NIH).

**Preparation of sonicated α-syn PFFs.** α-syn PFFs were prepared according to the previous methods reported by Volpicelli-Daley *et al*[21]. Mouse recombinant full-length α-syn was cloned into the ampicillin-resistant bacterial expression vector pRK172. The plasmids were then transformed into BL21(DE3)RIL-competent E. coli (Cat#: 230245; Life technologies). After bacterial growth, α-syn monomers were purified through several steps including anion exchange, dialysis, and size exclusion chromatography following the instructions of the aforementioned literature. *In vitro* α-syn fibrils were assembled by agitation in an Eppendorf orbital mixer (Cat#: 538400020) under 7 days of 37 $^{o}$C incubation at 1,000 rpm. Short fragments of α-syn fibrils were achieved simply by sonication at 20% amplitude for a total of 60 pulses (~0.5 seconds each) using sonicator with 1/8” probe-sonicator.

**Biotinylation of GQDs and Binding assay.** 50 mg of GQDs were dissolved in conjugation buffer (pH 4.7) and 12.5 mg of EDC reagent – N-(3-Dimethylaminopropyl)-N’-ethylcarbodiimide hydrochloride – (Cat#: 03449; Sigma-Aldrich) were subsequently added to replace the carboxyl groups. After 1 hour of reaction with vigorous stirring, 25 mg of EZ-Link

Amine-PEG$_3$-Biotin (Cat#: 21347; Thermo Scientific) were added to EDC-activated GQDs. The solution was then subjected to dialysis and rotary evaporation to yield the final product in powder form. For the binding assay between GQDs and α-syn fibrils, 5 mg/ml of α-syn fibrils were incubated with 5 mg/ml of biotinylated GQDs and streptavidin conjugated 0.8 nm ultra-small gold particles (Cat#: 800.099; Aurion) for 1 hour. Next, the streptavidin conjugated ultra-small gold particles bound with high affinity to biotinylated-GQDs were enhanced with GoldEnhance™ EM Plus solution (Cat#: 2114; Nanoprobes) for 5 mins. Non-reacted solution was removed by 100 kDa MWCO spin column (Cat#: UFC510024; Millipore-Sigma), followed by TEM analysis.

**Purification of $^{15}$N-lablled α-syn.** α-syn gene cloned in pRK172 vector was transformed into Escherichia coli BL21 (DE3) for α-syn overexpression. For the preparation of isotope-labelled α-syn, cells were grown in M9 minimal medium containing 0.5 g of $^{15}NH_4Cl$ and 1 g of $^{13}C$ glucose (Cambridge Isotope Laboratory Inc., Andover, MA) per liter at 37 °C with 100 μg/ml ampicillin. After the induction with isopropyl β–D-1-thiogalactopyranoside (IPTG), the heat-treated cell lysate was subjected to successive purifications using DEAE-Sephacel anion-exchange, Sephacryl S-200 size-exclusion, and S-Sepharose cation-exchange chromatography. The purified α-syn was dialysed against 12 L of fresh 20 mM 2-(N-morpholino)ethanesulfonic acid (MES) buffer (pH 6.5) three times, and stored in aliquots at a concentration of 1 mg/ml at -80 °C. The sample was concentrated to 5 mg/ml using Nanosep 10 K membrane (Pall Gelman, Germany) at 4 °C right before the experiments.

**NMR spectroscopy analysis**. A detailed NMR study of the interaction between α-syn and GQDs was performed using a 950 MHz spectrometer equipped with a cryo-genic probe (Bruker, Germany). $^{15}$N-lablled α-syn (5 mg/ml, 100 μl) was reacted with GQDs (5 mg/ml, 100 μl) under 37 °C shaking incubation at 1,000 rpm for 3 days. $^{15}$N-labelled α-syn samples were prepared using 20 mM MES buffer (pH 6.5) containing 10% for $^{1}$H-$^{15}$N HSQC measurements. $^{1}$H-$^{15}$N HSQC spectra of both α-syn and GQDs-reacted α-syn were acquired at 37 °C. Obtained data were processed by NMRPipe[22] and analysed by Sparky[23].

**Simulation details.** 200 ns molecular dynamics (MD) simulation was performed with Gromacs 5.1 to examine the interaction between GQDs and α-syn fibrils at molecular level[24]. The initial structure of hydrophobic NAC domain (residue 71 to 82) of α-syn was adapted from the ssNMR structure (PDB ID: 2N0A) with CHARMM forcefield[25] and the structure of GQDs was designed with CGenFF[26] by the protocols of https://cgenff.paramchem.org.

**CD measurements**. α-syn fibrils (5 mg/ml, 100 μl) were mixed with GQDs solution (5 mg/ml, 100 μl), and depolymerised under 37 °C shaking incubation at 1,000 rpm for 7 days. For far-UV CD measurements, the sample was diluted (1/2) with DW. The CD spectra between 190 and 260 nm were measured at 0.5 nm-interval using a J-815 spectropolarimeter (Jasco, Japan) and a 0.2 mm-path length quartz cuvette. The spectrum of the buffer solution was subtracted from the sample spectra. CD signals were normalised to the mean residue ellipticity, [$\theta$] with the unit of deg cm$^2$/dmol. The fractional secondary structure contents of α-syn fibrils and depolymerised α-syn fibrils were calculated using the algorithm of CONTIN/LL on the DichroWeb online server (http://dichroweb.cryst.bbk.ac.uk). For the calculation using CONTIN/LL, reference set 7 on Dichroweb was employed, which is optimised for the wavelength range of 190 to 240 nm.

**Primary neuron culture.** Primary cortical neurons were cultured using embryonic day 15 C57BL/6 mice (Charles River). Isolated neurons were plated onto the poly-*D*-lysine-coated (50 μg/ml) dishes (Cat#: P6407; Sigma-Aldrich) and placed under the culture medium comprised of Neurobasal Media (Cat#: 21103049; Life technologies) with B27 supplement (Cat#: 17504044;

Life technologies) and L-glutamine (Cat#: 25030149; Life technologies). The cultures were incubated in a 37 $^{o}$C, 7% $CO_2$ incubator and undergone a half medium change twice a week. To prevent glial cell growth, 30 μM 5-fluoro-2'-deocyuridine (Cat#: F0503; Sigma-Aldrich) was added after 5 days of culture. All procedures involving mice were approved by and conformed to the guidelines of the Johns Hopkins University Animal Care and Use Committee.

**Cell viability and Cytotoxicity assays.** For cell viability and cytotoxicity assays, primary cortical neurons were plated at a density 10,000 cells/cm$^2$ onto poly-*D*-lysine coated glass coverslip and incubated in 7% $CO_2$ incubator at 37 $^{o}$C with a half medium change twice a week. Cytotoxicity of primary cultured neurons was determined with α-syn PFFs (1 μg/ml) in the absence and presence of GQDs (1 μg/ml) in 10 DIV mouse cortical neurons for 7 days by LDH cytotoxicity assay kit (Cat#: 88954, Pierce), following the manufacturer's instructions. Apoptotic cell death was determined using TUNEL assay kit (Cat#: 12156792910; Roche). Primary cultured neuronal viability was quantified by alamarBlue cell viability assay kit (Cat#: DAL1025; Molecular Probes™) and neurite outgrowth staining kit (Cat#: A15001; Molecular Probes™), following the manufacturer's instructions.

***In vitro* immunofluorescence.** The mouse primary cortical neurons at a density of 20,000 cells/cm$^2$ were plated onto poly-*D*-lysine-coated coverslips. To fix the neurons, 4% PFA was used and blocked for 1 hour in a PBS solution containing 5% normal donkey serum (Cat#: 017-000-121; Jackson ImmunoResearch), 2% bovine serum albumin (Cat#: A7030; Sigma-Aldrich) and 0.1% Triton X-100 (Cat#: T8787; Sigma-Aldrich) at RT. Subsequent incubations with anti-8-OHG (Cat#: ab62623; 1:1,000; abcam), anti-pS129-α-syn (Cat#: ab59264; 1:1,000; abcam) and anti-MAP2 (Cat#: MAB3418; 1:1,000; Millipore) antibodies were carried out overnight at 4 $^{o}$C. After washing the samples with 0.1% Triton X-100 in PBS, the coverslips were incubated with a mixture of Cy3-conjugated (Donkey anti-mouse CY3; Cat#: 715-165-151, Donkey anti-rabbit CY3; Cat#: 711-165-152, Jackson ImmunoResearch) and FITC-conjugated (Donkey anti-mouse FITC; Cat#: 715-095-151, Donkey anti-rabbit FITC; Cat#: 711-095-152, Jackson ImmunoResearch) secondary antibodies for 1 hour at RT. The fluorescence images were taken through a Zeiss confocal microscope (LSM 710, Zeiss Confocal).

**Microfluidic chambers.** Prior to be affixed to the microfluidic devices (triple-chamber microfluidic devices, Cat#: TCND1000, Xona), the coverslips were prepared according to the previous method[5]. Each chamber was plated with approximately 100,000 neurons. At DIV 7, 0.5 μg of GQDs were added to chamber 1 (C1) or chamber 2 (C2) before the treatment of 0.5 μg of α-syn PFFs to C1. A 50 μl-difference in media volume was controlled through three compartments to regulate the direction of flow. 14 days post-α-syn PFFs treatment, neurons were fixed using 4% PFA in PBS. The fixed neurons in chambers were blocked in PBS solution containing 5% normal donkey serum, 2% bovine serum albumin and 0.1% Triton X-100 for 1 hour at RT. Neurons were subsequently incubated with anti-pS129-α-syn (Cat#: ab59264; 1:1,000; abcam) and anti-MAP2 (Cat#: MAB3418; 1:1,000; Millipore) antibodies for overnight at 4 $^{o}$C. After washing the chambers with 0.1% Triton X-100 in PBS, they were incubated in a mixture of FITC-conjugated (Jackson ImmunoResearch) and Cy3-conjugated (Jackson ImmunoResearch) secondary antibodies for 1 hour at RT. The fluorescent images were acquired through a Zeiss confocal microscope

**Animals.** All experimental procedures were followed according to the guidelines of Laboratory Animal Manual of the National Institute of Health Guide to the Care and Use of Animals, which were approved by the Johns Hopkins Medical Institute Animal Care and Use Committee. The human a α-syn-A53T transgenic mice (B6.Cg-Tg; Prnp-SNCA*A53T;23Mkle/J, stock#: 006823) were purchased at the Jackson Lab[27].

**Stereological assessments.** All experimental procedures were followed according to the guidelines of Laboratory Animal Manual of the National Institute of Health Guide to the Care and Use of Animals, which were approved by the Johns Hopkins Medical Institute Animal Care and Use Committee. C57BL6 mice (8-10 week-old male) were purchased from the Jackson laboratories. After anesthetising with pentobarbital (60 mg/kg), mice were stereotaxically injected in a designated coordinate of the striatum (+2.0 mm from midline, +2.6 mm beneath the dura, +0.2 mm relative to Bregma) with of PBS (2 μl) or PFFs (5 μg/2 μl) (Cat#: Model 900; David KOPF instruments). As a treatment, 50 μl of GQDs (50 μg per mouse) were i.p. injected for 6 months on a biweekly basis. After 6 months, animals were perfused with PBS and fixed by 4% PFA for 12 hours. The brains were then subjected to cryoprotection with 30% sucrose. For immunohistochemistry, 50 μm coronal sections were cut throughout the brain including the SN. Every 4$^{th}$ section was used for analysis and incubated with the rabbit polyclonal anti-TH (Cat#: NB300-19; 1:1,000; Novus Biologicals), rabbit polyclonal anti-pS129-α-syn (Cat#: ab59264; 1:1,000; abcam) with blocking solution. The signals were visualised using DAB kit (Cat#: SK-4100; Vector Laboratories) followed by incubation with streptavidin-conjugated horseradish peroxidase and biotinylated secondary antibodies (Cat#: PK-6101; Vector Laboratories). To identify the Nissl substance, the TH-stained tissues were prepared and counterstained with thionin on slides. Total number of TH- and Nissl-positive neurons in the SN were counted using an Optical Factionator probe of Stereo Investigator software (MBF Bioscience).

***In vivo* immunohistochemistry.** The brains of animals were perfused with PBS followed by 4% PFA. After post-fixation and cryoprotection, sections were stained with anti-Iba-1 (Cat#: 019-19741; 1:1,000; Wako) or anti-GFAP (Cat#: Z0334; 1:2,000; Dako) antibodies followed by incubation with biotin-conjugated anti-rabbit antibody and ABC reagents (Cat#: PK-6101; Vector Laboratories). The sections were developed using DAB peroxidase substrate (Cat#: SK-4100; Vector Laboratories). The number of microglia and densities of astrocytes in the SN were measured with ImageJ software (http://rsb.info.nih.gov/ij/, NIH). For histopathology of the major organs, 8-10 week-old male C57BL/6 mice were i.p. injected with 50 μg of GQDs on a biweekly basis for 6 months. After 6 months of injection period, animals were perfused with PBS followed 4% PFA. The liver, kidney, and spleen were isolated and stained with H&E staining kit (Cat#: H-3502; Vector Laboratories).

**Behaviour analyses.** The cylinder test[28]: The test was devised to identify any asymmetry in forelimb uses. A blinded observer scored every contact made by a forepaw in a 20 cm-wide clear glass cylinder from the video clips. A total number of 20 – 30 wall touches were counted per animal (only the contacts with fully extended forelimbs). The degree of impaired forelimb contacts was quantified as a percentage based on the total forelimb uses. Control group scores approximately 50% in this test. The animals were completely isolated from the test cylinder prior to the actual testing. No habituation of the animals to the testing cylinder was allowed before video recording.

Pole test[29]: Test animals were allowed to adapt to the behavioural room for 30 mins before actual testing. A 75 cm metal pole with a diameter of 9 mm was prepared with bandage gauze wrapping. To initiate the test, each mouse was placed 7.5 cm from the top facing upwards. The scores were based on the total time taken to reach the bottom. The test animals were trained for two successive days before the actual trial where each practice provided three different trials. No more than 60 secs was allowed for each animal. Results for the turn down, climb down, and the total time (in sec) were recorded.

**Statistics.** Data were presented as mean ± SD from at least three independent experiments. In order to assess the statistical significance, Student's t tests or ANOVA tests followed by Bonferroni post hoc analysis were performed using Prism6 software (GraphPad). Assessments with a $p < 0.05$ were considered significant.

**Data Availability.** There is no data relevant to accession codes or unique identifiers non-publicly available. The data that support the plots within this paper and other findings of this study are available from the corresponding authors upon reasonable request.

# Figure Legends

**Figure 1. The effect of GQDs on α-syn fibrillization and fibril disaggregation. a,** Schematic representation of the α-syn fibrillization (5 mg/ml α-syn monomers) and disaggregation (5 mg/ml α-syn fibrils) in the presence and absence of GQDs (5 mg/ml). **b,** The kinetics of α-syn fibrillization using aliquots of reaction monitored by ThT fluorescence and **c,** turbidity assays (n=4, biologically independent samples; two-way ANOVA with a post hoc Bonferroni test, ***$P < 0.001$; error bars are the standard deviation). The mean values and p values are given in Supplementary Table 1. **d,** TEM images of α-syn after fibrillization in the absence (left) and presence (right) of GQDs. **e,** The kinetics of preformed α-syn fibrils after incubation with GQDs, aliquots of reaction monitored by ThT fluorescence and **f,** turbidity assays at various time points (n=4, biologically independent samples; two-way ANOVA with a post hoc Bonferroni test, ***$P < 0.001$; error bars are the standard deviation). The mean values and p values are given in Supplementary Table 1. **g,** Quantifications of the end-to-end length and number of α-syn fibrils. Mean values of end-to-end length are 937.84, 245.52, 123.13, 66.27, and 51.02 at 0, 6, 12, 24, and 72 hours (n=50 fibrils at each time point; one-way ANOVA with a post hoc Bonferroni test, $P$ values are $P < 0.0001$, $P < 0.0001$, $P < 0.0001$, and $P < 0.0001$ at 6, 12, 24, and 72 hours; error bars are the standard deviation). Mean values of number are 15.53, 37.01, 63.83, 97.52, and 44.38 at 0, 6, 12, 24, and 72 hours (n=6, biologically independent samples; one-way ANOVA with a post hoc Bonferroni test, $P$ values are $P = 0.0483$, $P < 0.0001$, $P < 0.0001$, and $P = 0.0050$ at 6, 12, 24, and 72 hours; error bars are the standard deviation). **h,** the amount of remaining α-syn fibrils determined by multiplying the end-to-end length and number of α-syn fibrils at the same time points (0, 6, 12, 24, and 72hours) in the presence of GQDs. Mean values are 100.00, 47.33, 42.00, 36.67, and 13.83 at 0, 6, 12, 24, and 72 hours (n=6, biologically independent

samples; one-way ANOVA with a post hoc Bonferroni test, error bars are standard deviation). **i,** TEM images of preformed α-syn fibrils after various time points (6, 12 hours, 1, 3, and 7days) in the absence (top) and presence (bottom) of GQDs. **j,** Representative image of α-syn fibrils by dot-blot assay at various time points (0, 12 hours, 1, 3, and 7 days) with α-syn filament specific antibody. These experiments were independently repeated three times with similar results. **k,** BN-PAGE analysis of α-syn, prepared with aliquots of reaction run after various time points (0, 3, 6 and 12 hours, 1, 3, and 7days). These experiments were independently repeated three times with similar results.

**Figure 2. Detailed analysis of the interaction between GQDs and mature α-syn fibril during the dissociation process. a,** TEM images for binding between biotinylated GQDs and α-syn fibrils with low and high magnifications. Single-headed arrows indicate the biotinylated GQDs enhanced with ultra-small gold-streptavidin nanoparticles. **b,** Quantifications of the average width of α-syn fibrils during the disaggregation process after 1 hour of incubation. Mean ± s.d. values are 14.74 ± 0.70 and 17.42 ± 0.83 for fibrils and fibrils+GQDs-biotin (n=20 for each group, two-tailed Student's t-test). **c**, NMR chemical shift difference obtained from full $^{1}H$-$^{15}N$ HSQC spectra. The decreased intensity ratio of NMR chemical shifts is presented for each residue after binding with GQDs. **d,** The time course simulation dynamics of the interaction between GQDs and mature α-syn fibril and 65 ns snapshot image (bottom right) of the interaction between GQDs and mature α-syn fibril with designated sidechains attributed for the major binding force. **e,** Time-dependent plots for RMSD of atomic positions, SASA for α-syn fibril only group and α-syn fibril and GQDs group, total potential energy ($\Delta U_{tot}$), electrostatic energy ($\Delta E_{elec}$), and van der Waals energy ($\Delta E_{van}$) for α-syn fibril and GQDs group, respectively

from the top. **f,** Time-dependent secondary structure plot calculated by the DSSP algorithm. **g,** CD spectra of α-syn monomers and α-syn fibrils without and with GQDs after 7 days of incubation. **h,** The fractional secondary structure contents ratio of α-syn fibrils and disaggregated α-syn fibrils by GQDs calculated using the algorithm of CONTIN/LL (n=5, biologically independent samples). Data are mean ± s. d. (fibrils vs fibrils+GQDs; Coil: 20.13 ± 5.74 vs 24.58 ± 3.64; β-Turn: 22.39 ± 4.62 vs 25.89 ± 1.34, β-Sheet: 53.30 ± 3.53 vs 29.76 ± 3.35; α-Helix: 4.17 ± 1.25 vs 19.74 ± 1.52).

**Figure 3. The effect of GQDs on α-syn PFFs-induced neuronal death, pathology and transmission *in vitro*. a.** Neuronal death assessed by TUNEL, **b,** alamarBlue and **c,** LDH assays treated with α-syn PFFs (1 μg/ml) in the absence and presence of GQDs (1 μg/ml) in 10 DIV mouse cortical neurons for 7 days. Mean values of TUNEL are 18.94, 15.86, 61.39, and 30.45 for PBS, GQDs, PFFs, and PFFs+GQDs; mean values of alamarBlue are 100.00, 99.16, 49.33, and 78.86 for PBS, GQDs, PFFs, and PFFs+GQDs; mean values of LDH assay are 100.00, 95.43, 154.58, and 113.33 for PBS, GQDs, PFFs, and PFFs+GQDs (n=6, biologically independent samples; two-way ANOVA with post hoc Bonferroni test; NS, not significant; error bars are the standard deviation). **d,** Representative immunoblot levels with p-α-syn antibody. **e,** Quantifications of the SDS-insoluble fraction normalised to the levels of β-actin. Mean values are 1.00, 1.49, 16.35, and 6.86 for PBS, GQDs, PFFs, and PFFs+GQDs (n=6, biologically independent samples; two-way ANOVA with post hoc Bonferroni test; NS, not significant; error bars are the standard deviation). **f,** Representative p-α-syn immunostaining micrographs with p-α-syn antibody. **g,** Quantifications of p-α-syn immunofluorescence intensities normalised to the PBS control. Mean values are 1.00, 0.99, 13.62, and 4.69 for PBS, GQDs, PFFs, and PFFs+GQDs (n=6, biologically independent samples; two-way ANOVA with post hoc Bonferroni test; NS, not significant; error bars are the standard deviation). **h,** Schematic representation of the microfluidic device for the transmission of pathologic α-syn, composed of three connected chambers. **i,** Representative images of p-α-syn immunostained neurons in the microfluidic device after 14 days post α-syn PFFs addition. **j,** Quantifications of p-α-syn immunofluorescence intensities. The areas occupied by p-α-syn were measured in each chamber. Mean values of chamber 1 are 15.67, 3.12, and 4.52 for PFFs, PFFs+GQDs (C1), and PFFs+GQDs (C2); mean values of chamber 2 are 8.52, 2.56, and 3.90 for PFFs, PFFs+GQDs (C1), and PFFs+GQDs (C2); mean values of chamber 3 are 6.04, 1.72, and 1.67 for PFFs, PFFs+GQDs (C1), and PFFs+GQDs (C2) (n=6, biologically independent samples; two-way ANOVA with post hoc Bonferroni test; NS, not significant; error bars are the standard deviation).

**Figure 4. The effect of GQDs on α-syn-induced pathologies *in vivo*. a,** Schematic illustration of injection coordinates of α-syn PFFs (5 μg) for stereotaxic intra-striatal injection in C57BL/6 mice. As a treatment, 50 μg of GQDs or PBS were i.p. injected biweekly for 6 months. **b,** Representative TH-immunohistochemistry images in the SN of α-syn PFFs-injected hemisphere in the absence (top) and the presence (bottom) of GQDs. **c,** Stereological counting of the number of TH- and Nissl-positive neurons in the SN via unbiased stereological analysis after 6 months of α-syn PFFs injection with and without GQDs injection. Mean values are 5061, 5096, 3155, and 4080 for Nissl-positive neurons; 4039, 4068, 2327, and 3221 for TH-positive neurons (n=6, biologically independent animals; two-way ANOVA with a post hoc Bonferroni test; NS, not significant; error bars are the standard deviation). **d,** Representative TH-immunohistochemistry images in the striatum of α-syn PFFs-injected hemisphere. **e,** Quantifications of TH-

immunopositive fibre densities in the striatum. Mean values are 1.00, 1.03, 0.42, and 0.90 for PBS+PBS, PBS+GQDs, PFFs+PBS, and PFFs+GQDs (n=6, biologically independent animals; two-way ANOVA with a post hoc Bonferroni test; NS, not significant; error bars are the standard deviation). **f,** Assessments of the behavioural deficits measured by the use of forepaws in the cylinder test. Mean values are 51.00, 51.67, 29.83, and 44.17 for PBS+PBS, PBS+GQDs, PFFs+PBS, and PFFs+GQDs (n=6, biologically independent animals; two-way ANOVA with a post hoc Bonferroni test; NS, not significant; error bars are the standard deviation). **g,** Assessments of the behavioural deficits measured by the ability to grasp and descend from a pole. Mean values are 10.25, 11.38, 22.73, and 14.43 for PBS+PBS, PBS+GQDs, PFFs+PBS, and PFFs+GQDs (n=6, biologically independent animals; two-way ANOVA with a post hoc Bonferroni test; NS, not significant; error bars are the standard deviation). **h,** Representative p-α-syn immunostaining images in the striatum and SN of α-syn PFFs-injected hemisphere. **i,** Quantifications of p-α-syn immunoreactive neurons in the striatum and SN. Mean values are 43.33 and 12.50 for PFFs+PBS and PFFs+GQDs of striatum (STR); 17.67 and 8.00 for PFFs+PBS and PFFs+GQDs of SN (n=6, biologically independent animals; two-tailed Student's t-test; error bars are the standard deviation). **j,** Distribution of LB/LN-like pathology in the CNS of α-syn PFFs-injected hemisphere (p-α-syn positive neurons; red dots, p-α-syn positive neurites; red lines).

# Figure 1

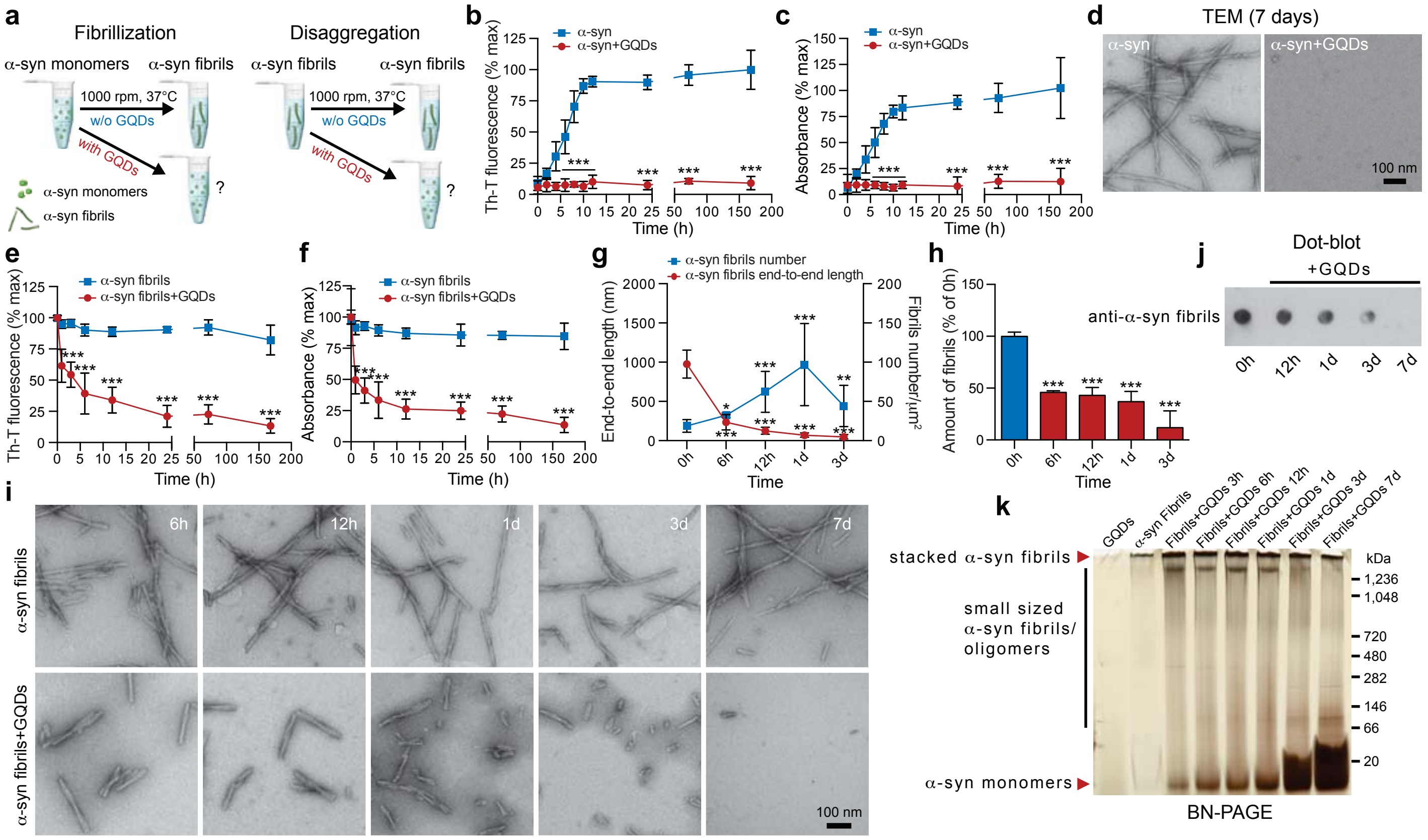

# Figure 2

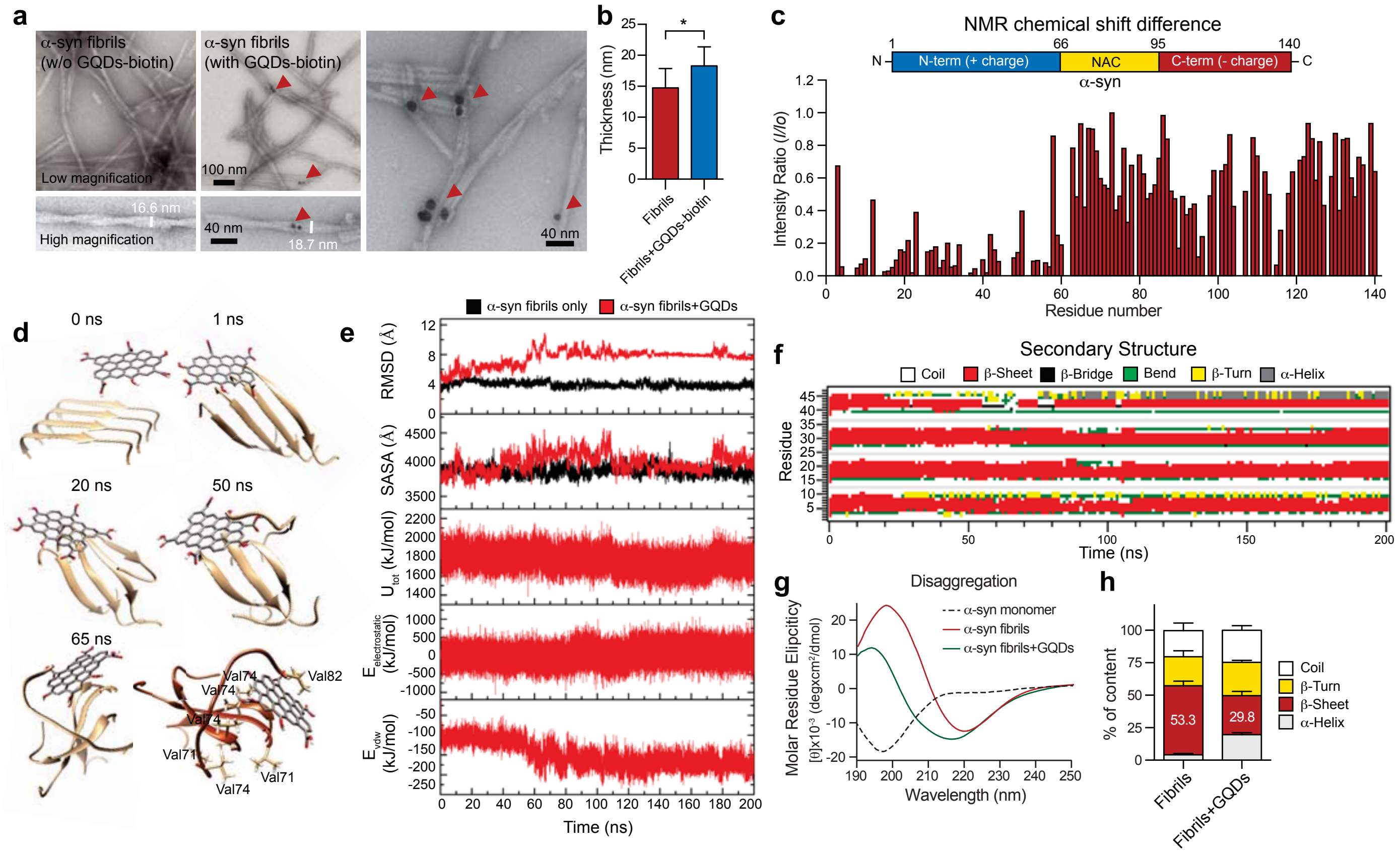

# Figure 3

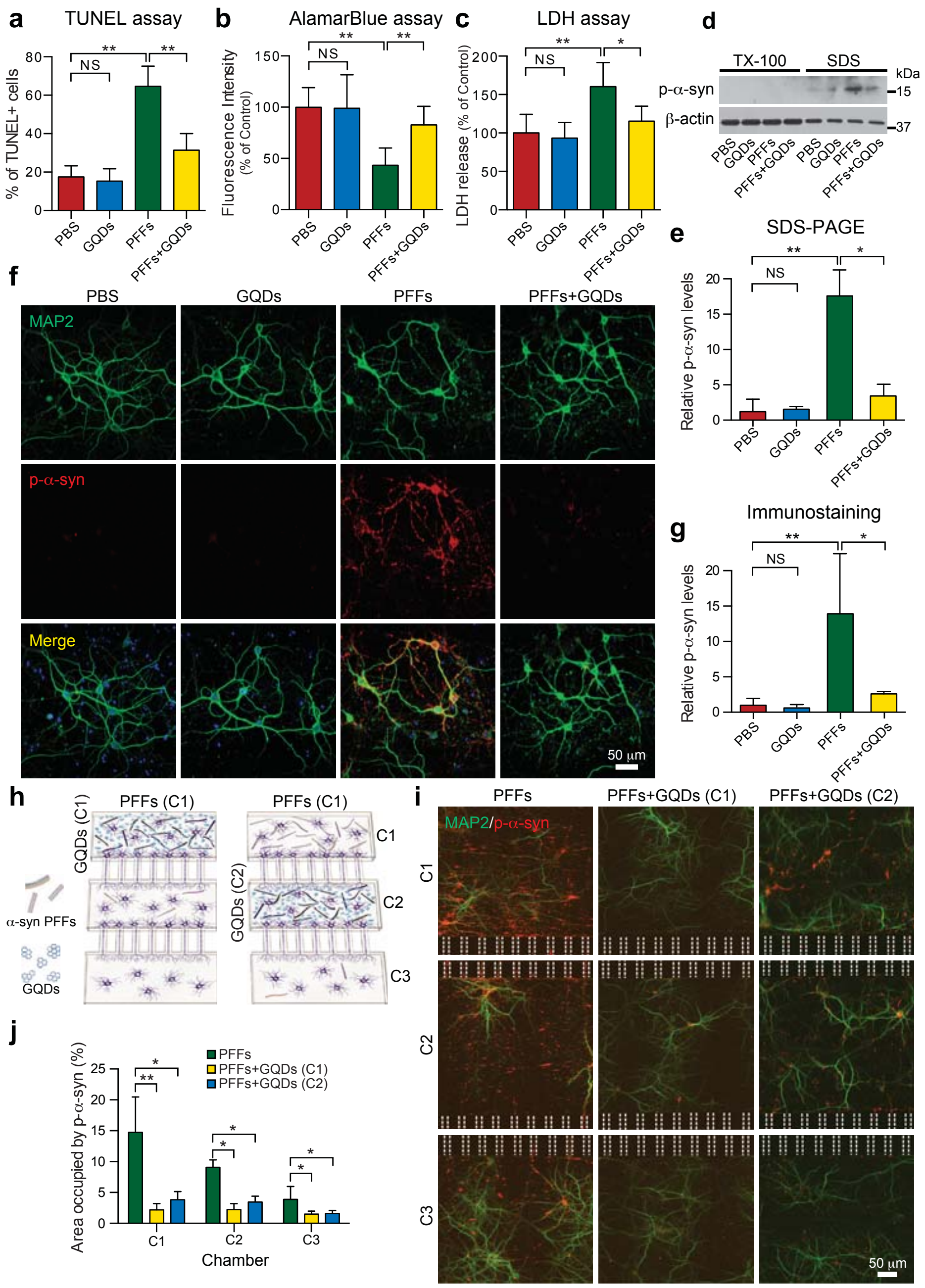

# Figure 4

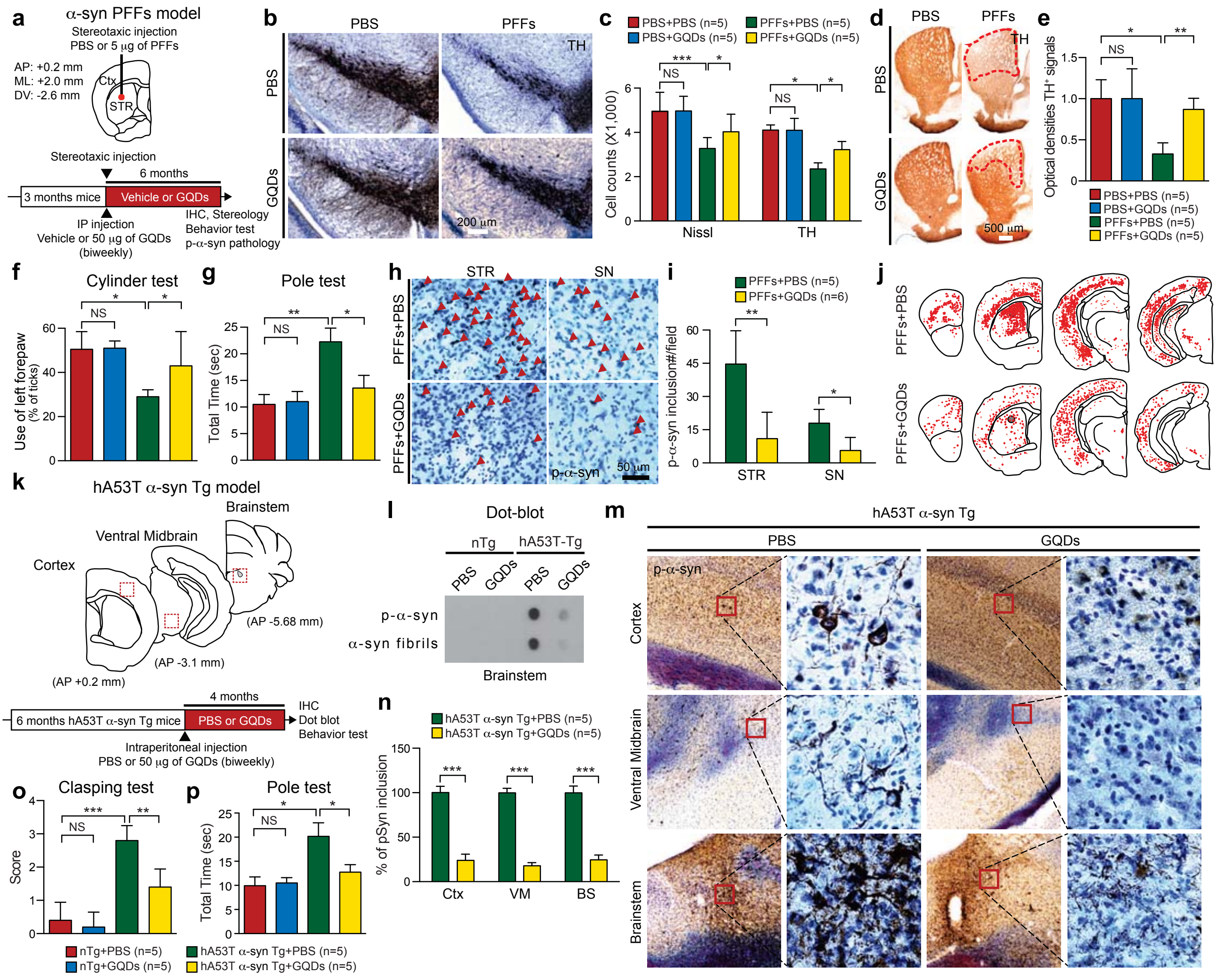

# Supplementary Figure 1

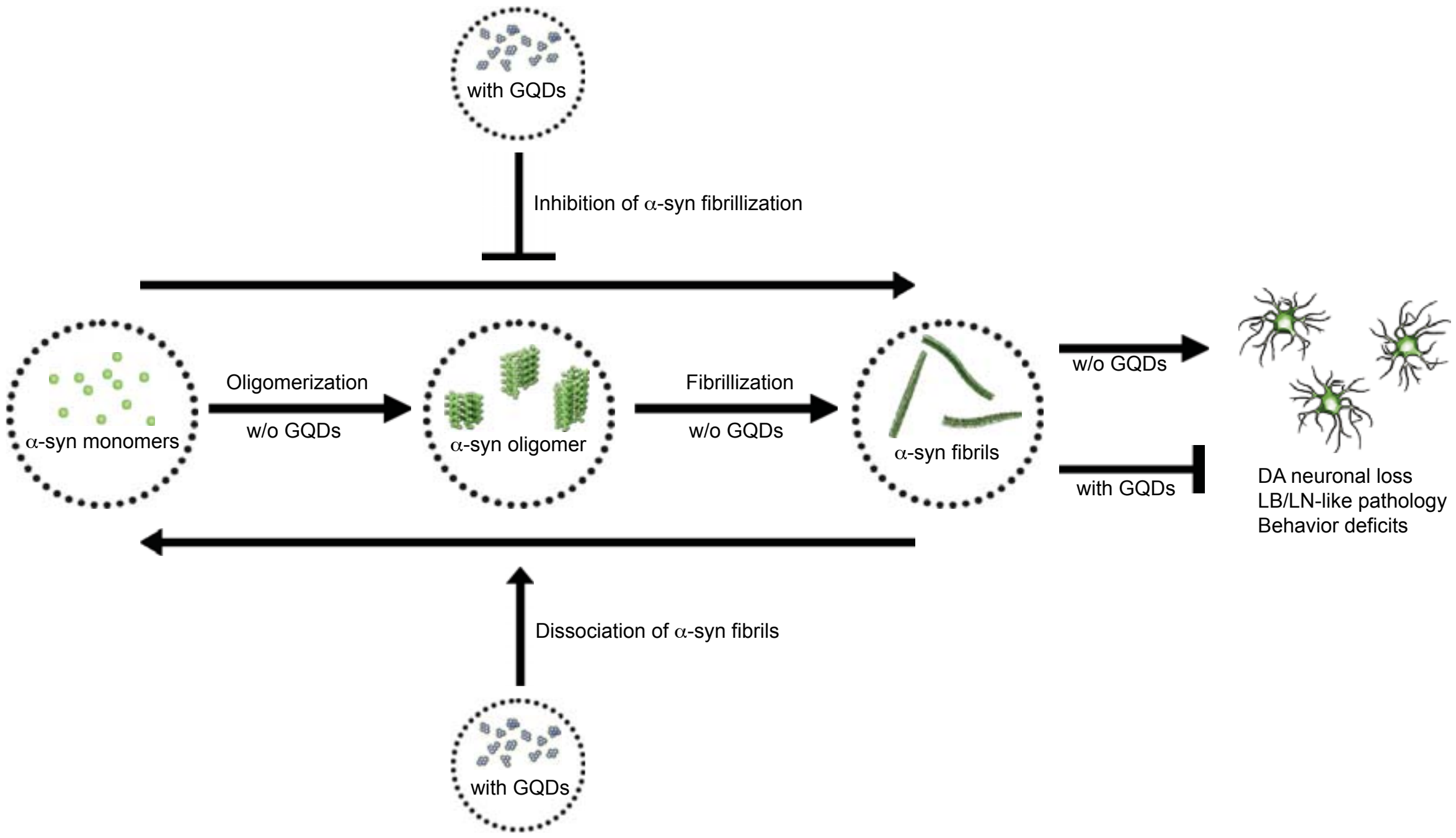

# Supplementary Figure 2

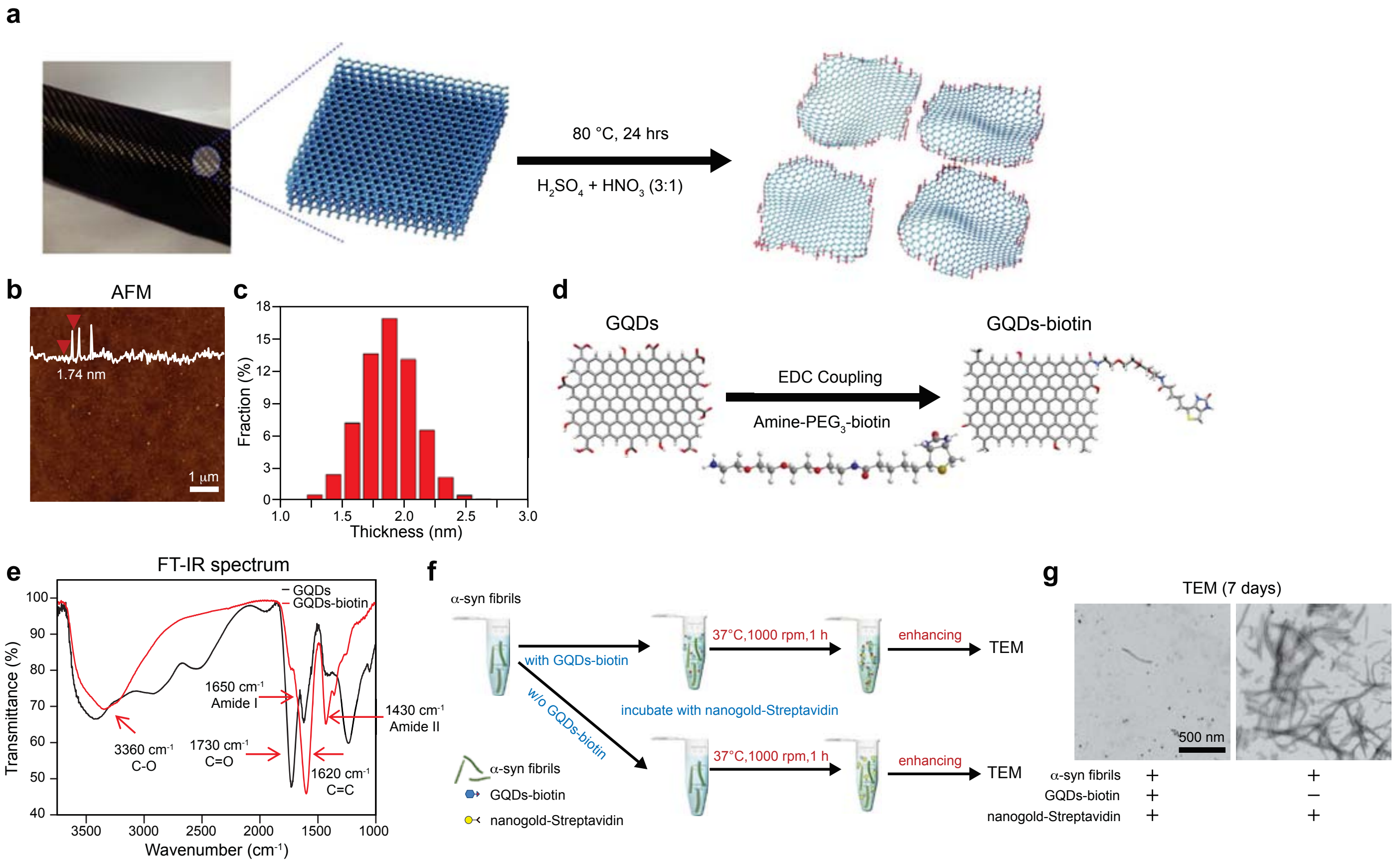

# Supplementary Figure 3

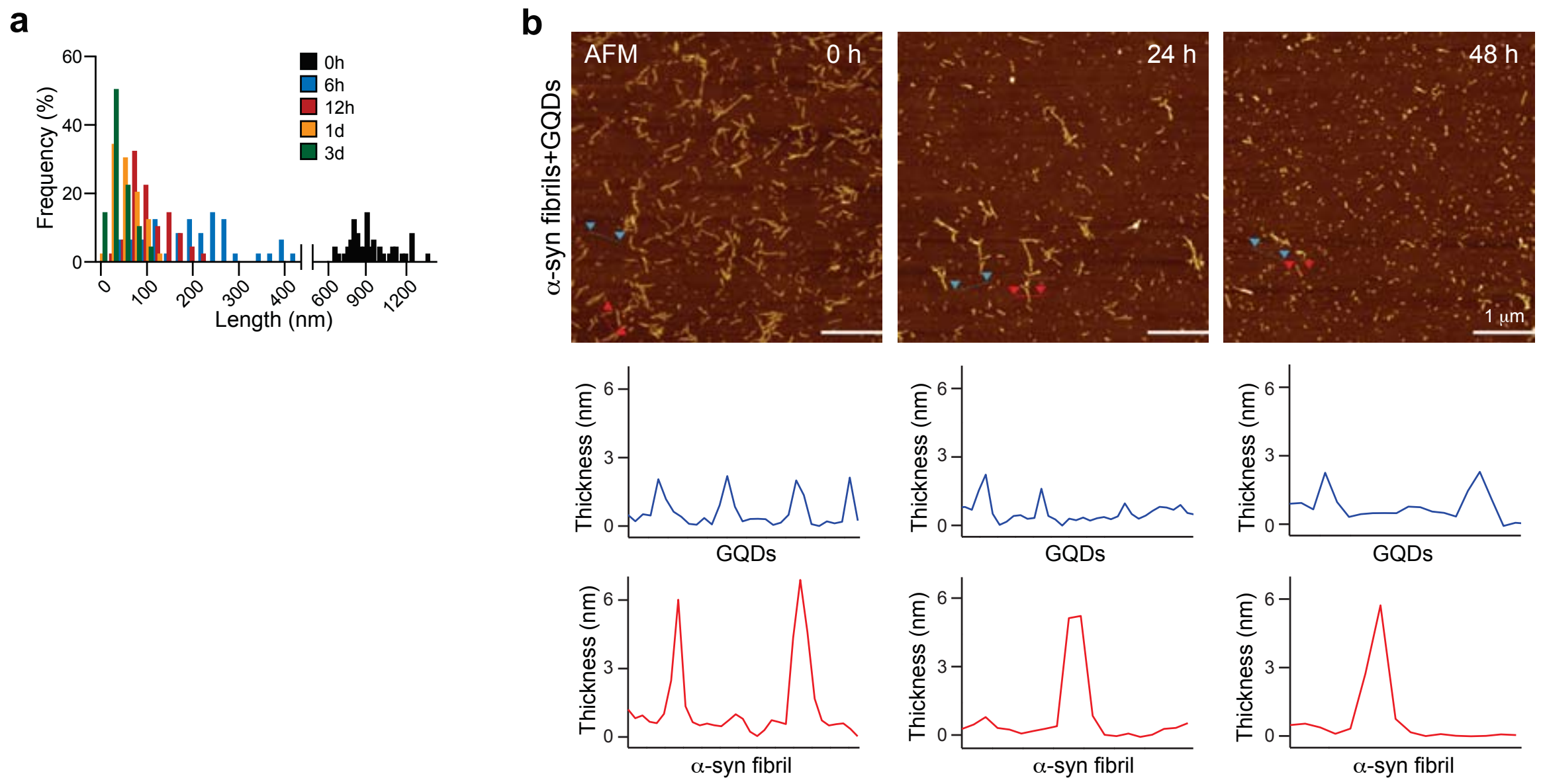

# Supplementary Figure 4

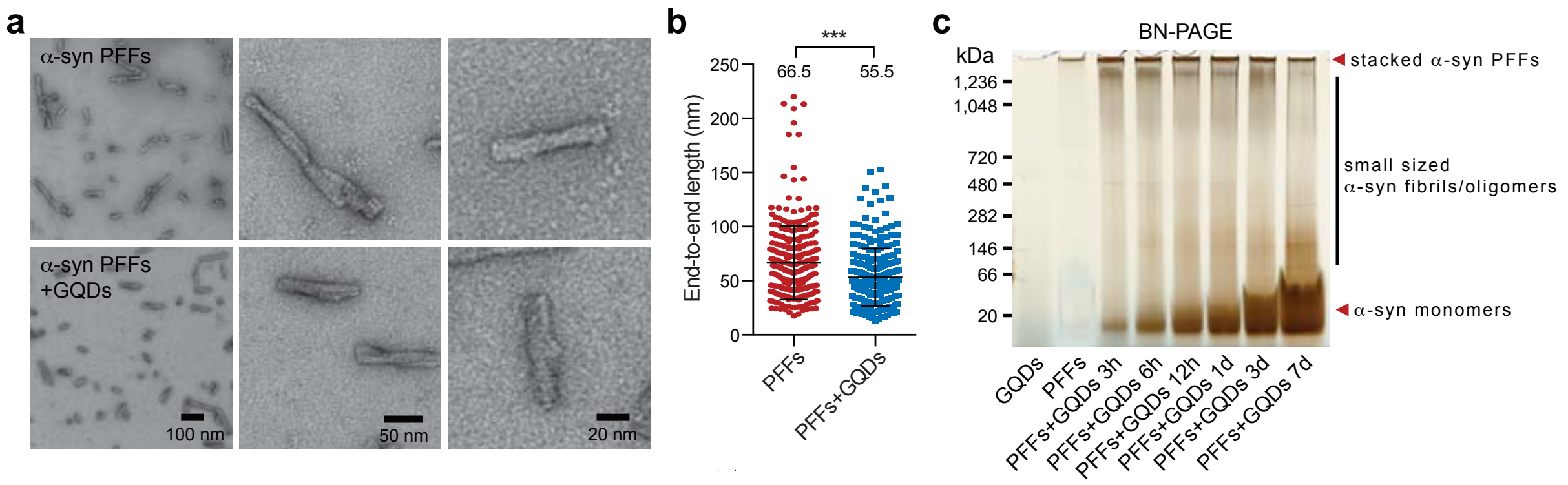

# Supplementary Figure 5

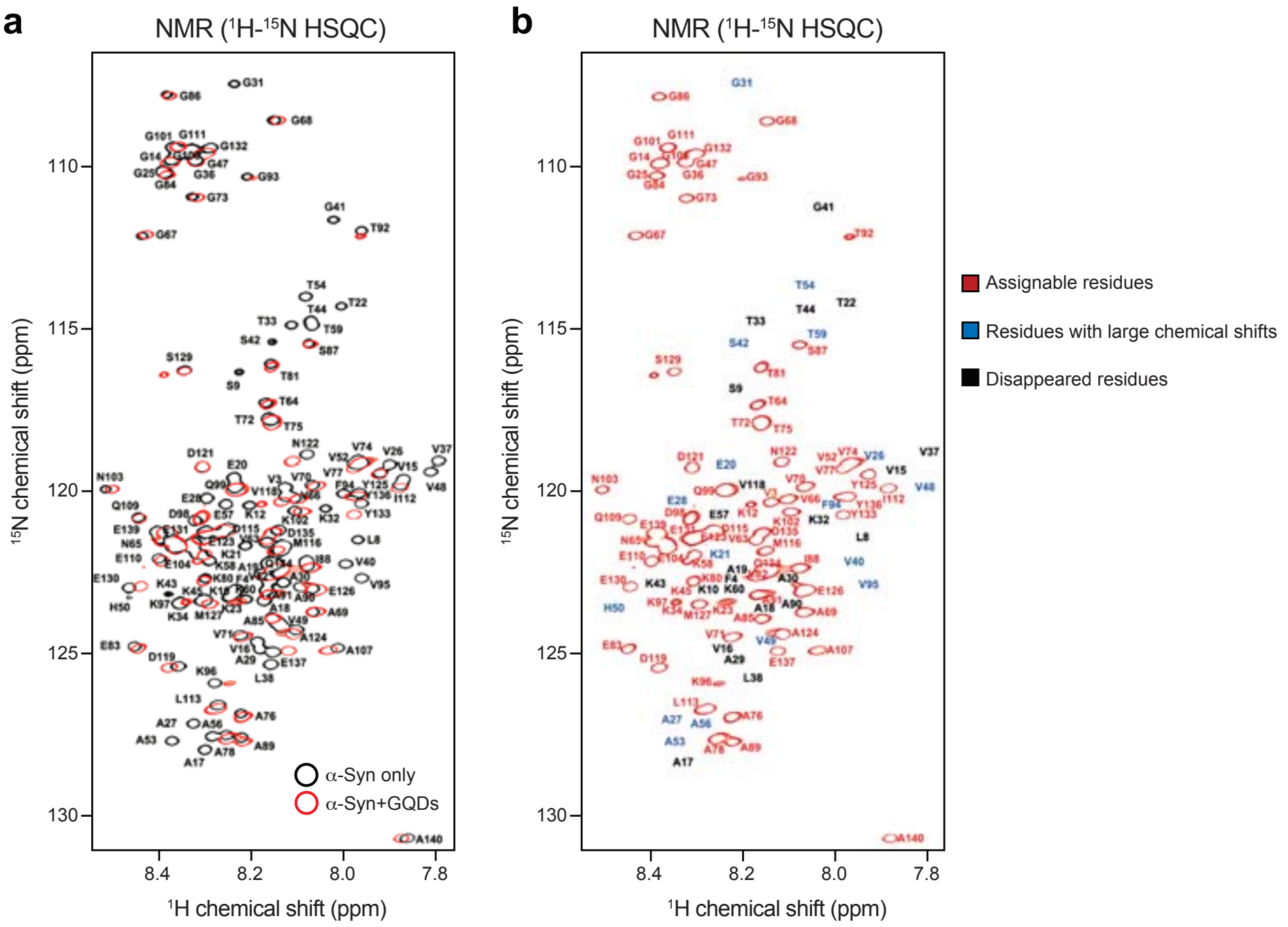

# Supplementary Figure 6

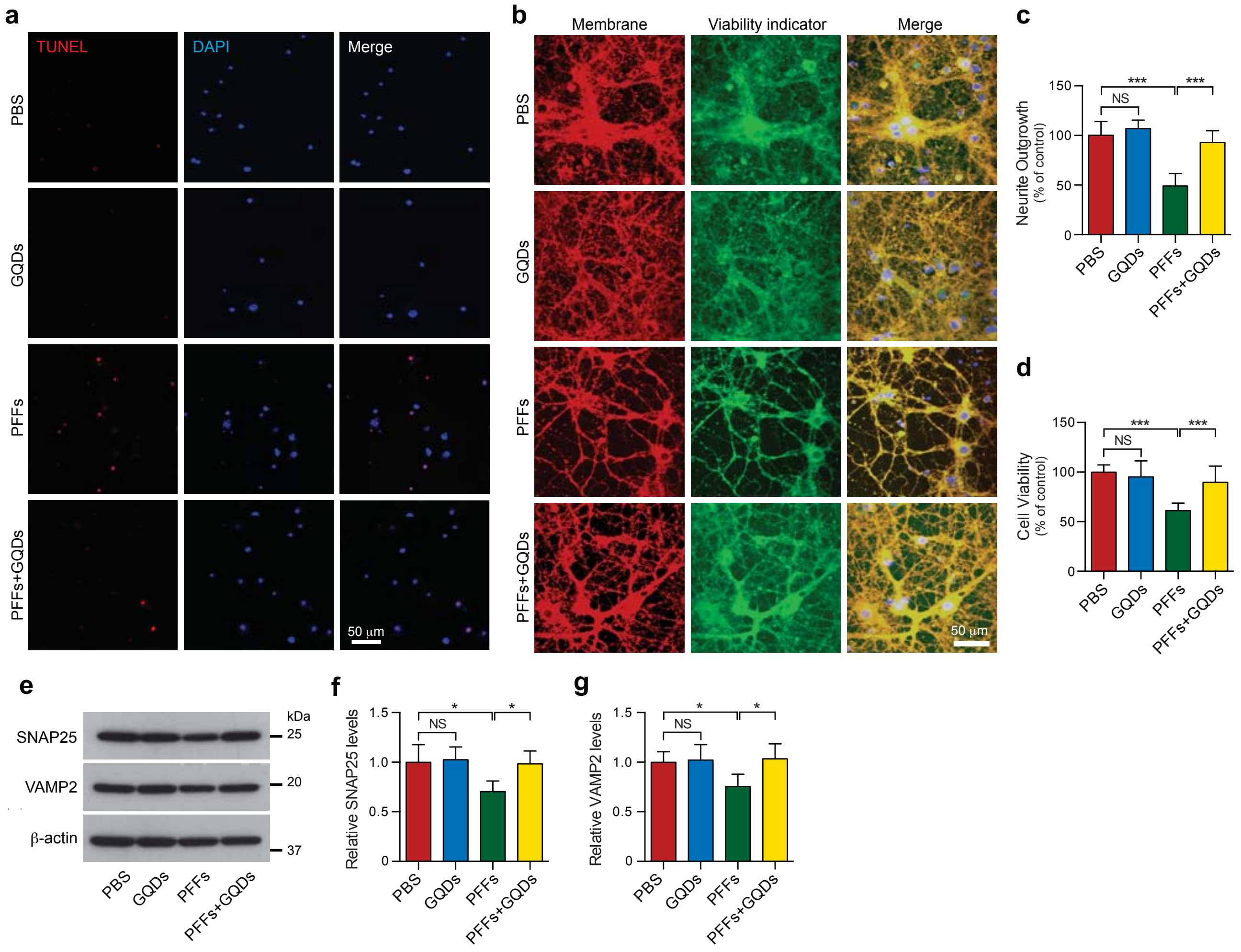

# Supplementary Figure 7

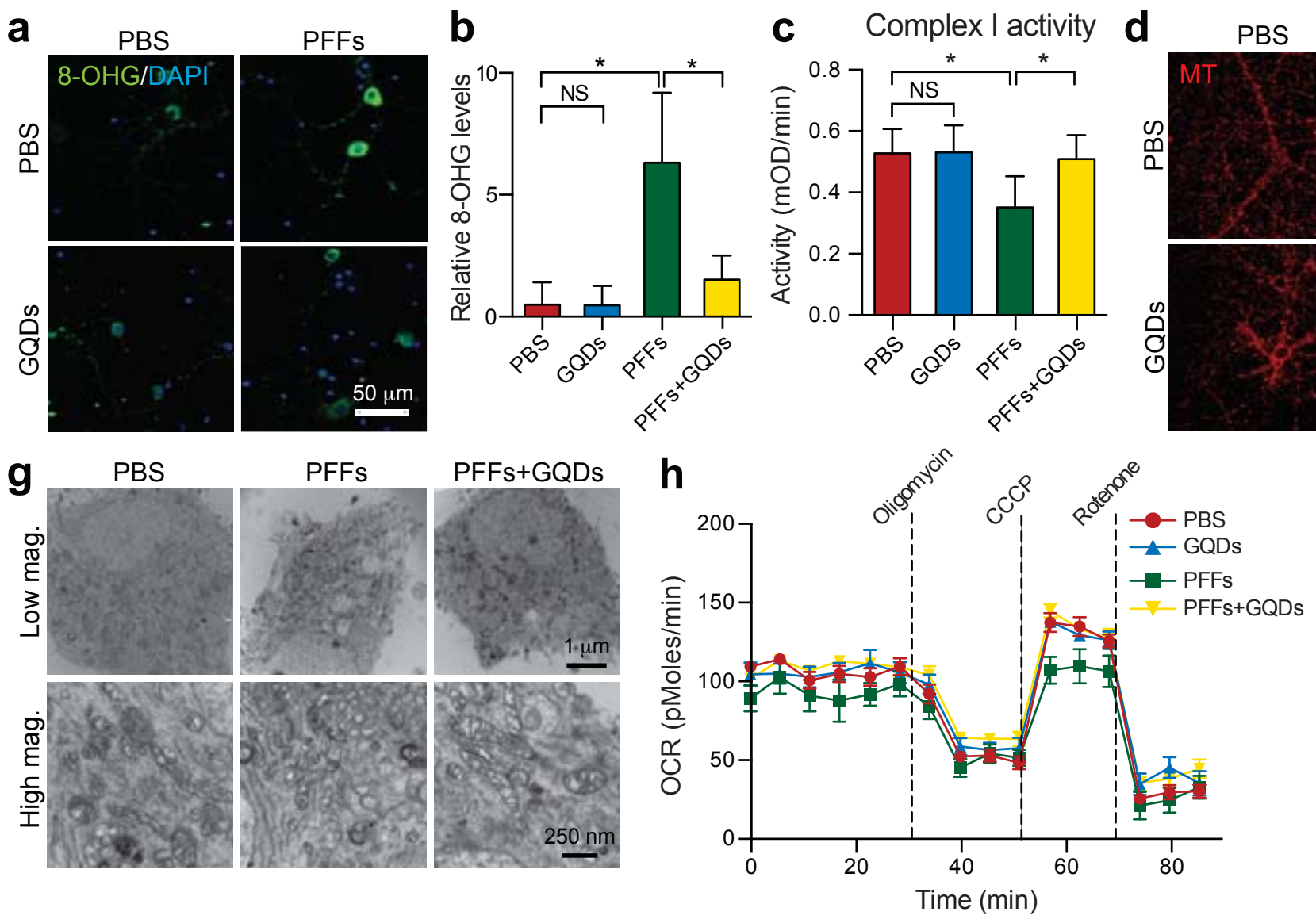

# Supplementary Figure 8

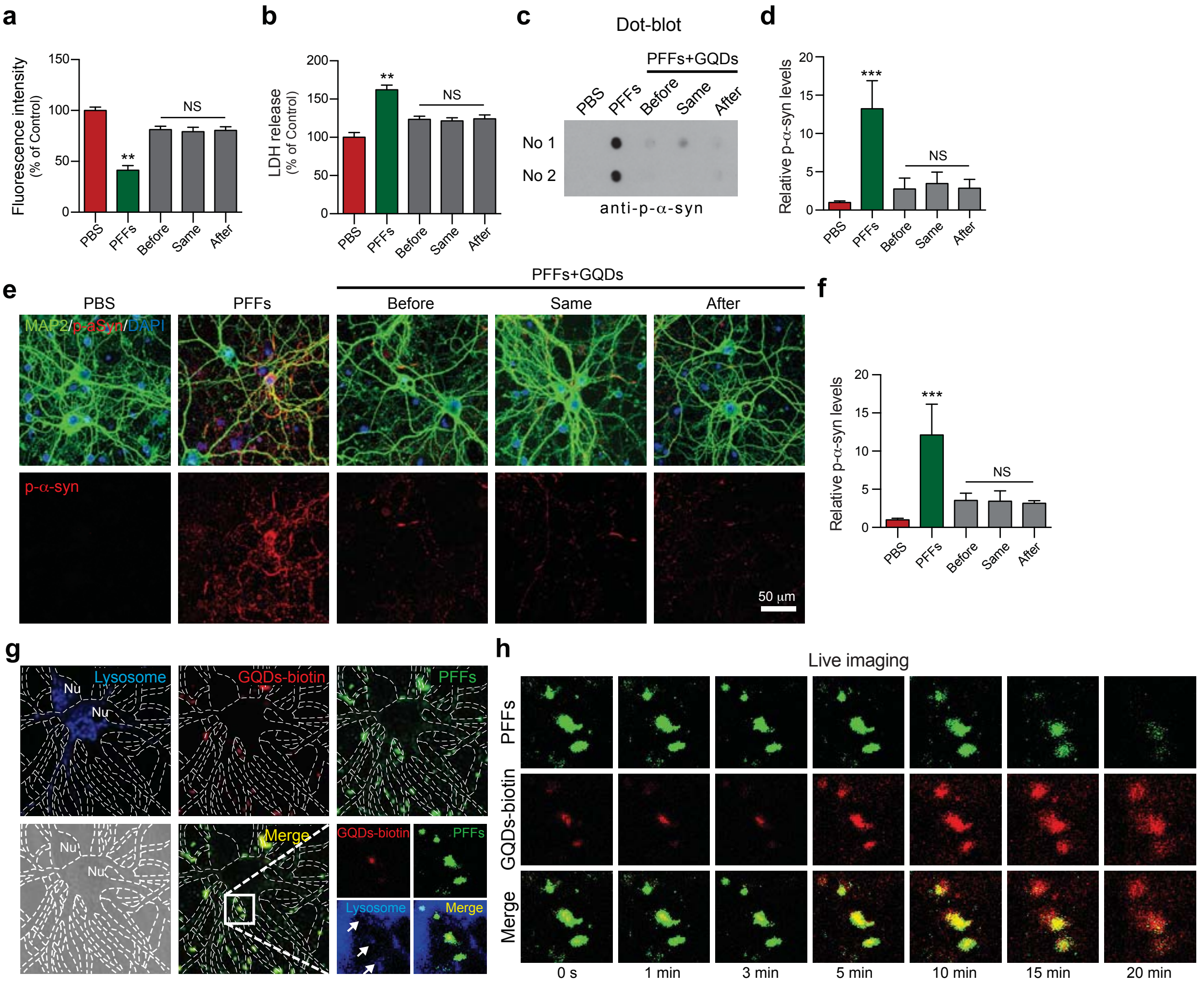

# Supplementary Figure 9

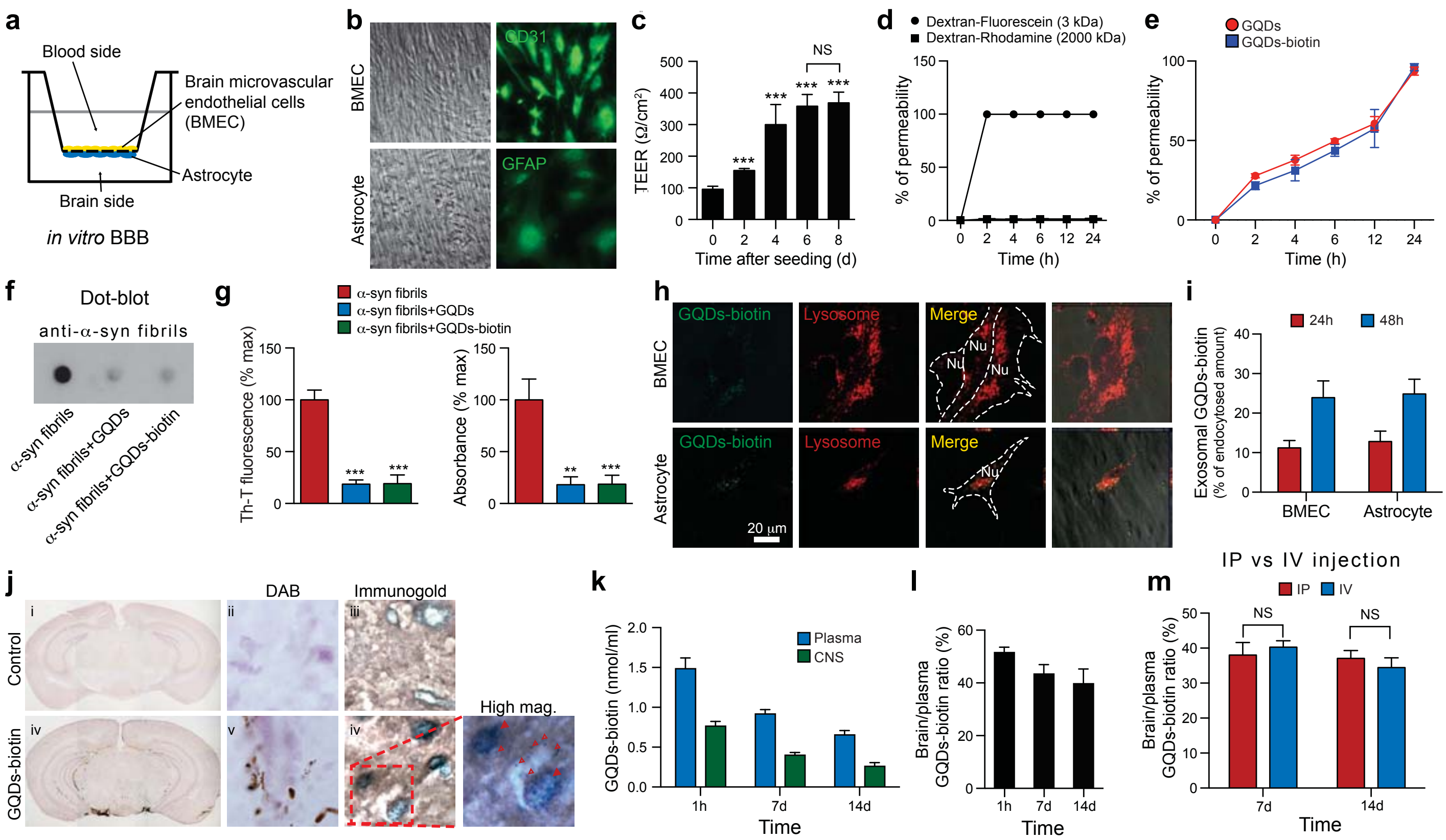

# Supplementary Figure 10

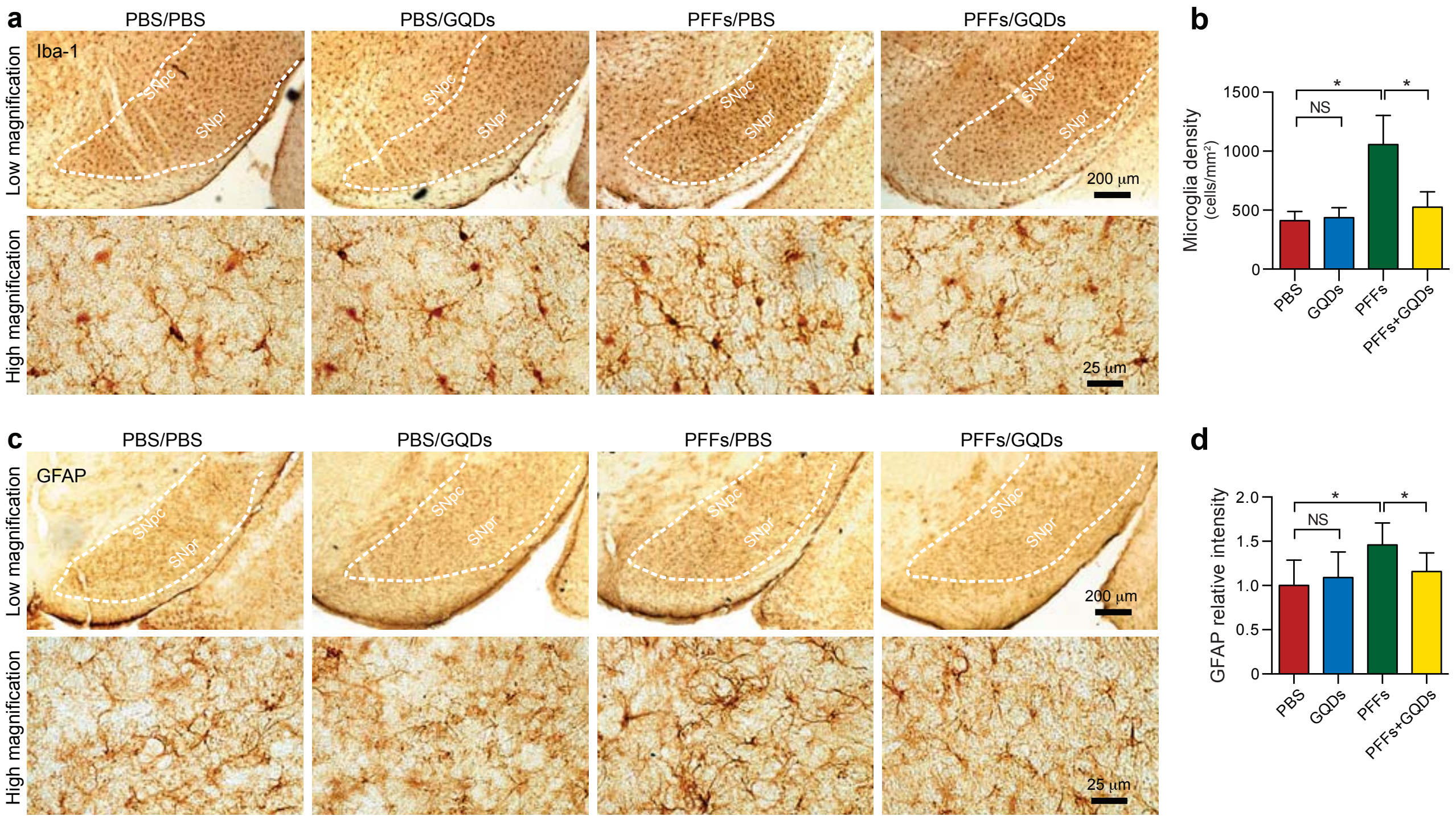

# Supplementary Figure 11

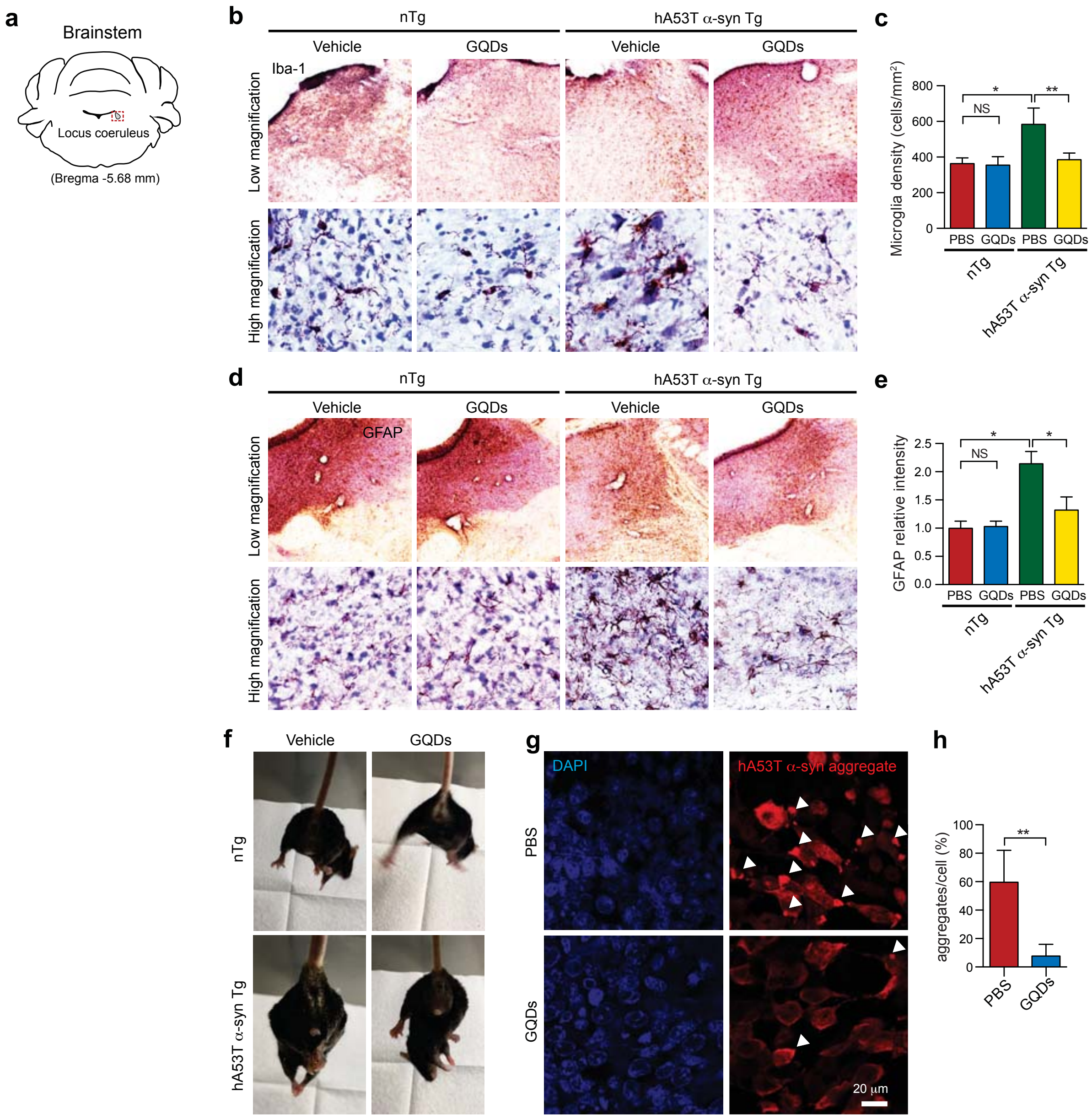

# Supplementary Figure 12

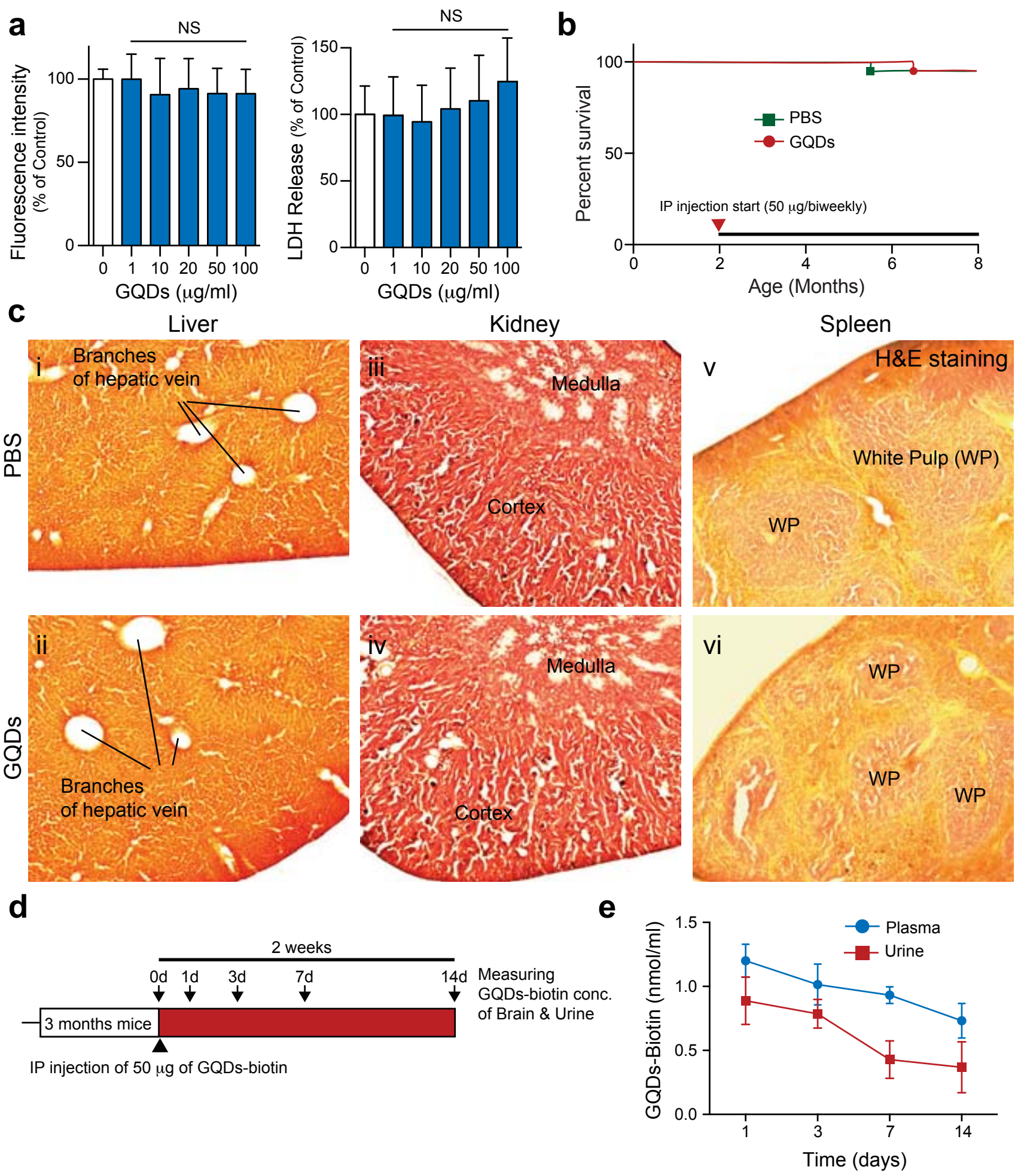

# Supplementary Figure 13

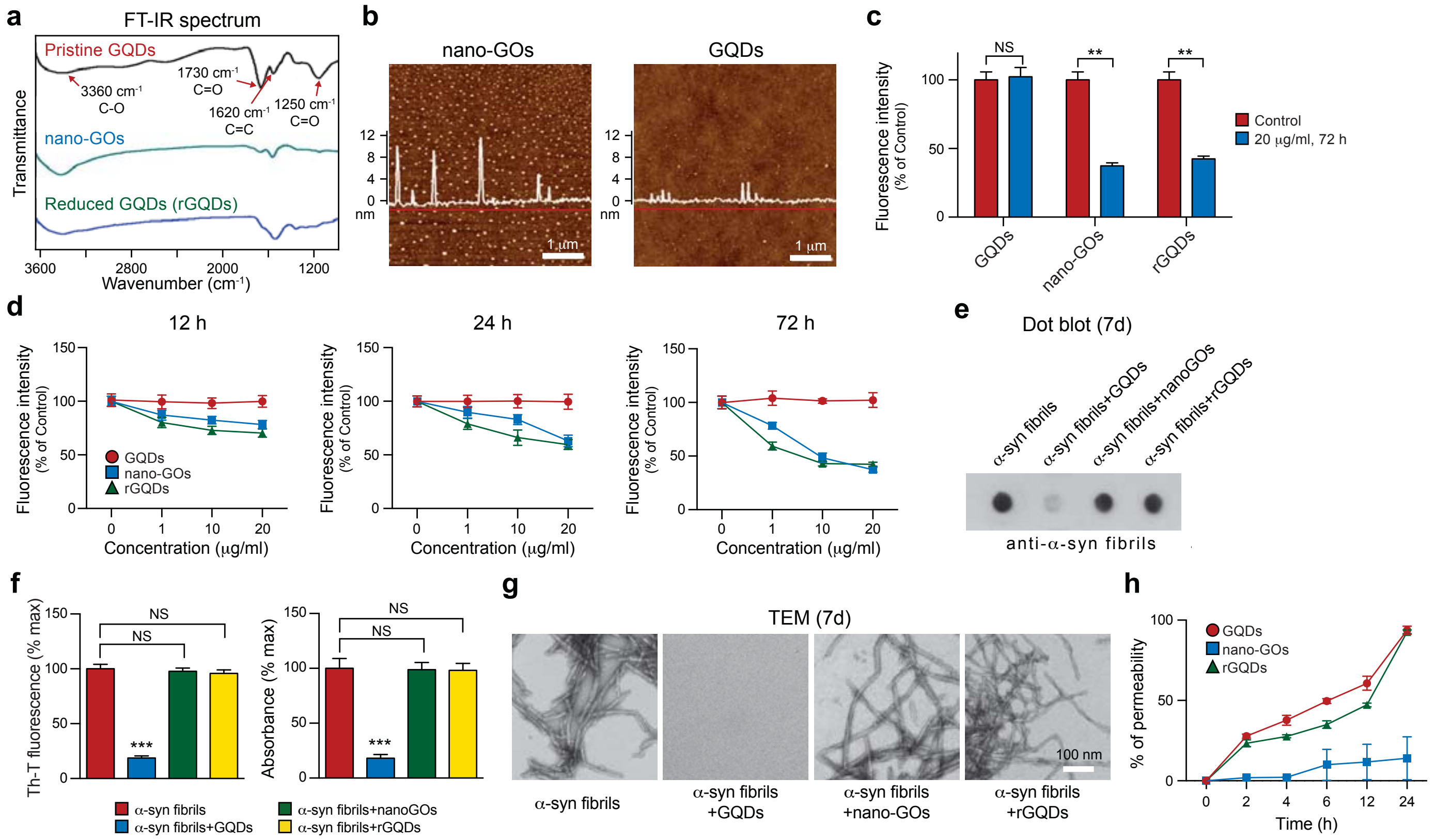